\documentclass[12pt]{article}
\pdfoutput=1

\usepackage{graphicx}
\usepackage{amsmath,amsfonts,amssymb,amsthm,latexsym}
\usepackage{graphicx}
\usepackage{bm}
\usepackage{mathtools}
\usepackage{natbib}
\usepackage{ifthen}
\usepackage{mathtools}
\usepackage[hidelinks]{hyperref}
\usepackage{bm}
\usepackage{xfrac}
\usepackage[dvipsnames]{xcolor}
\usepackage[margin=1.25in]{geometry}
\usepackage{setspace}
\usepackage[noabbrev,capitalize]{cleveref}
\usepackage{makecell}
\usepackage{tikz}
\usetikzlibrary{positioning}
\usetikzlibrary{arrows}

\usepackage{pgfplots}
\usetikzlibrary{patterns}
\usepgfplotslibrary{fillbetween}

\newcommand{\agind}[1][i]{_{#1}}

\newcommand{\ironed}{\bar}
\newcommand{\constrained}{\hat}
\newcommand{\optconstrained}{\composed{\optimized}{\constrained}}
\newcommand{\optimized}[1]{#1\opt}
\newcommand{\differentiated}[1]{#1'}

\newcommand{\tagged}[2]{{#2}^{#1}}

\newcommand{\primedarg}[1]{#1\primed}
\newcommand{\noaccents}[1]{#1}
\newcommand{\composed}[3]{#1{#2{#3}}}

\newcommand{\newagentvar}[3][\noaccents]{%
\expandafter\newcommand\expandafter{\csname #2\endcsname}{#1{#3}}%
\expandafter\newcommand\expandafter{\csname #2s\endcsname}{#1{\boldsymbol{#3}}}%
\expandafter\newcommand\expandafter{\csname #2smi\endcsname}[1][i]{#1{\boldsymbol{#3}}_{-##1}}%
\expandafter\newcommand\expandafter{\csname #2i\endcsname}[1][i]{#1{#3}\agind[##1]}%
\expandafter\newcommand\expandafter{\csname #2ith\endcsname}[1][i]{#1{#3}_{(##1)}}%
}

\newcommand{\newitemvar}[3][\noaccents]{%
\expandafter\newcommand\expandafter{\csname #2\endcsname}{#1{#3}}%
\expandafter\newcommand\expandafter{\csname #2s\endcsname}{#1{\boldsymbol{#3}}}%
\expandafter\newcommand\expandafter{\csname #2smj\endcsname}[1][j]{#1{\boldsymbol{#3}}_{-##1}}%
\expandafter\newcommand\expandafter{\csname #2j\endcsname}[1][j]{#1{#3}_{##1}}%
\expandafter\newcommand\expandafter{\csname #2jth\endcsname}[1][j]{#1{#3}_{(##1)}}%
}

\newcommand{\forrezs}[1]{{#1}^{\rezs}}
\newagentvar[\forrezs]{imbal}{\phi}
\newagentvar[\forrezs]{cumimbal}{\Phi}

\newagentvar{alloc}{x}

\newagentvar{falloc}{z}

\newagentvar{quant}{q}
\newagentvar[\constrained]{exquant}{\quant}
\newagentvar[\constrained]{critquant}{\quant}  
\newagentvar{Val}{\nu}
\newagentvar{toquant}{Q}

\newagentvar{qprice}{\price}
\newagentvar{qrev}{R}
\newagentvar[\ironed]{iqrev}{\qrev}

\newagentvar[\bar]{bstrat}{\strat}
\newagentvar[\bar]{bpay}{\pay}
\newagentvar[\bar]{bwal}{\wal}
\newagentvar[\bar]{bbid}{\bid}
\newagentvar{zpay}{\pay^{\rezs}}
\newagentvar{zwal}{\wal^{\rezs}}
\newagentvar{zdelt}{\Delta^{\rezs}}
\newagentvar{bomega}{\bm{\omega}}

\newagentvar{qalloc}{y}
\newagentvar{cumalloc}{Y}
\newagentvar[\constrained]{calloc}{\qalloc}
\newagentvar[\constrained]{cumcalloc}{\cumalloc}
\newagentvar[\ironed]{ialloc}{\qalloc}
\newagentvar[\ironed]{icumalloc}{\cumalloc}

\newcommand{\exposted}[1]{#1^{\text{\it EP}}}
\newagentvar[\composed{\exposted}{\constrained}]{excalloc}{\qalloc}
\newagentvar[\exposted]{exalloc}{\qalloc}
\newagentvar[\exposted]{extalloc}{\talloc}
\newagentvar[\exposted]{exfalloc}{\falloc}

\newagentvar{rev}{R}
\newagentvar[\differentiated]{marg}{\rev}
\newagentvar{rawrev}{P}
\newagentvar[\differentiated]{rawmarg}{\rawrev}

\newagentvar[\primedarg]{inducedrev}{\rev}

\newagentvar[\tilde]{pseudorev}{\rev}
\newagentvar[\tilde]{pseudorawrev}{\rawrev}

\newagentvar{typespace}{{\cal T}}
\newagentvar{typesubspace}{S}

\newagentvar{type}{t}
\newagentvar{othertype}{s}
\newagentvar{val}{v}
\newagentvar{outcome}{w}
\newagentvar{outcomespace}{{\cal W}}

\newcommand{\served}[1]{#1^1}
\newcommand{\nonserved}[1]{#1^0}
\newcommand{\alloced}[1]{#1^{\alloc}}
\newcommand{\allocedi}[1]{#1^{\alloci}}
\newagentvar[\alloced]{xoutcome}{\outcome}
\newagentvar[\allocedi]{xioutcome}{\outcome}
\newagentvar[\served]{soutcome}{\outcome}

\newagentvar[\nonserved]{nsoutcome}{\outcome}

\newagentvar{price}{p}
\newagentvar{talloc}{\alloc}
\newagentvar[\tilde]{eval}{\val}
\newagentvar[\tilde]{balloc}{\alloc}
\newagentvar[\tilde]{dalloc}{y}
\newagentvar[\tilde]{eballoc}{y}
\newagentvar{ealloc}{y}
\newagentvar[\tilde]{bprice}{\price}

\newagentvar{act}{a}
\newagentvar{bidspace}{A}
\newagentvar{actspace}{A}

\newagentvar[\constrained]{critval}{\val}
\newagentvar[\constrained]{crittype}{\type}
\newagentvar[\constrained]{critvirt}{\virt}
\newagentvar[\constrained]{reserve}{\val} 
\newagentvar[\constrained]{bidreserve}{\bid} 
\newagentvar[\optconstrained]{monop}{\val}
\newagentvar[\constrained]{monot}{\type}
\newagentvar[\optimized]{monorev}{\rev}

\newagentvar{rcalloc}{y}
\newagentvar[\optimized]{optrcalloc}{\rcalloc}
\newagentvar{biddist}{G}

\newagentvar[\constrained]{critbid}{\bid}
\newagentvar[\constrained]{cbid}{B}
\newagentvar[\primedarg]{wbid}{\bid}
\newagentvar{gfunc}{\vartheta}
\newagentvar{block}{C}

\newagentvar{util}{u}

\newagentvar{strat}{\beta}
\newagentvar{bid}{b}

\newagentvar{pay}{\pi}
\newagentvar{rez}{\rho}
\newagentvar[\tilde]{trez}{\rez}

\newagentvar{virt}{\phi}
\newagentvar{cumvirt}{\Phi}
\newagentvar{qvirt}{\phi}
\newagentvar[\ironed]{ivirt}{\virt}
\newagentvar[\ironed]{icumvirt}{\cumvirt}

\newagentvar[\tagged{\text{SD}}]{sdvirt}{\virt}
\newagentvar[\composed{\tagged{\text{SD}}}{\ironed}]{sdivirt}{\virt}
\newagentvar[\tagged{\text{MD}}]{mdvirt}{\virt}
\newagentvar[\composed{\tagged{\text{MD}}}{\ironed}]{mdivirt}{\virt}

\newagentvar{dist}{F}
\newagentvar{dens}{f}
\newagentvar{hazard}{h}
\newagentvar{cumhazard}{H}

\newagentvar[\ironed]{iprice}{\price}  
\newagentvar[\ironed]{ival}{\val}  

\newagentvar{ints}{{\cal I}}

\newagentvar{wal}{w}
\newagentvar{Util}{U}

\newitemvar{pos}{j}
\newitemvar{weight}{w}
\newitemvar[\differentiated]{mweight}{\weight}
\newitemvar{udtype}{\type}
\newitemvar[\constrained]{udcrittype}{\type}
\newitemvar{udalloc}{\alloc}
\newitemvar{udprice}{\price}
\newitemvar[\constrained]{udcalloc}{\qalloc}
\newitemvar[\constrained]{udcumcalloc}{\cumalloc}

\newagentvar{mech}{M}
\newagentvar[\skew{5}{\hat}]{cmech}{{\cal M}}
\newagentvar{alg}{{\cal A}}

\DeclareMathOperator{\OPT}{OPT}

\DeclareMathOperator{\APX}{APX}

\newcommand{\mecha}{{M}}

\newagentvar{trans}{\sigma}

\newagentvar{demandset}{S}

\newcommand{\opt}{^{\star}}
\newcommand{\primed}{^\dagger}

\newagentvar{distout}{w}
\newagentvar{idistout}{\bar{w}}

\newagentvar{gap}{\delta}

\newcommand{\density}{f}

\newcommand{\valf}{V}
\newcommand{\ratio}{r}
\newcommand{\ScaledQuant}{\hat{Q}}
\DeclareMathOperator{\regu}{Reg}
\newcommand{\regular}{\feasibleDist^{\regu}}
\DeclareMathOperator{\tri}{Tri}
\DeclareMathOperator{\Trunc}{Trunc}
\newcommand{\TRIF}{\feasibleDist^{\tri}}
\newcommand{\TRUNCF}{\feasibleDist^{\Trunc}}
\newcommand{\QRF}{\feasibleDist^{\Qr}}
\DeclareMathOperator{\Qr}{Qr}

\DeclareMathOperator{\SMKUP}{SMKUP}

\newcommand{\stocmarkup}{\feasibleMecha^{\SMKUP}}

\newcommand{\distTri}{\dist^{\tri}}
\newcommand{\distTrunc}{\dist^{\Trunc}}
\newcommand{\distQr}{\dist^{\Qr}}

\newcommand{\vsec}{\val_{(2)}}

\newcommand{\mechaAR}{\mecha_{\alpha, r}}

\newcommand{\feasibleDist}{\mathcal{F}}

\newcommand{\feasibleMecha}{\mathcal{M}}
\newcommand{\qo}{\bar{\quant}}
\newcommand{\qt}{\bar{\quant}'\!}
\newcommand{\revq}{P_{\qt}}
\newcommand{\valq}{\valf_{\qt}}
\newcommand{\revqt}{P_{\qt'}}
\newcommand{\valqt}{\valf_{\qt'}}
\newcommand{\quantf}{Q}

\newcommand{\mechaARO}{\mecha_{\optweight, \optratio}}
\newcommand{\aval}{0.80564048}
\newcommand{\rval}{2.4469452}
\newcommand{\qval}{0.0931057}
\newcommand{\apxval}{1.9068943}
\newcommand{\asimp}{0.806}
\newcommand{\rsimp}{2.447}
\newcommand{\qsimp}{0.093}
\newcommand{\apxsimp}{1.907}

\newcommand{\secprob}{\alpha}
\newcommand{\optf}[1]{\OPT_{#1}(#1)}
\newcommand{\optweight}{\secprob^*}
\newcommand{\optratio}{\ratio^*}
\newcommand{\optqo}{\qo^*}

\newcommand{\piratio}{\beta}

\newcommand{\valscale}{\alpha}

\newcommand{\alphaL}{0.81}
\newcommand{\alphaS}{0.8}
\newcommand{\ratioL}{2.449}
\newcommand{\ratioS}{2.445}
\newcommand{\qoL}{0.094}
\newcommand{\qoS}{0.093}

\newcommand{\APXSPA}{\APX_1}
\newcommand{\APXR}{\APX_{*}}

\DeclareMathOperator{\MKUP}{MKUP}
\DeclareMathOperator{\SPA}{SPA}

\newcommand{\discreteQ}{Q_d}

\newcommand{\precision}{\epsilon}

\newcommand{\discreteR}{R_d}
\newcommand{\precisionR}{\precision_r}

\newcommand{\virtual}{\phi}
\newcommand{\pRev}{P}
\newtheorem{theorem}{Theorem}
\newtheorem{definition}{Definition}
\newtheorem{lemma}{Lemma}

\newtheorem{corollary}{Corollary}

\newtheorem{claim}{Claim}

\newtheorem{assumption}{Assumption}

\DeclareMathOperator*{\argmax}{argmax}
\DeclareMathOperator*{\argmin}{argmin}

\newcommand{\given}{\,\mid\,}

\newcommand{\prob}[2][]{\text{\bf Pr}\ifthenelse{\not\equal{}{#1}}{_{#1}}{}\!\left[{\def\givenn{\middle|}#2}\right]}
\newcommand{\expect}[2][]{\text{\bf E}\ifthenelse{\not\equal{}{#1}}{_{#1}}{}\!\left[{\def\givenn{\middle|}#2}\right]}

\newcommand{\tparen}{\big}
\newcommand{\tprob}[2][]{\text{\bf Pr}\ifthenelse{\not\equal{}{#1}}{_{#1}}{}\tparen[{\def\given{\tparen|}#2}\tparen]}
\newcommand{\texpect}[2][]{\text{\bf E}\ifthenelse{\not\equal{}{#1}}{_{#1}}{}\tparen[{\def\given{\tparen|}#2}\tparen]}

\newcommand{\sprob}[2][]{\text{\bf Pr}\ifthenelse{\not\equal{}{#1}}{_{#1}}{}[#2]}
\newcommand{\sexpect}[2][]{\text{\bf E}\ifthenelse{\not\equal{}{#1}}{_{#1}}{}[#2]}

 \crefname{assumption}{Assumption}{Assumptions}

\begin{document}




\providecommand{\keywords}[1]
{
  \small	
  \textbf{\textit{Keywords---}} #1
}

\providecommand{\JEL}[1]
{
  \small	
  \textbf{\textit{JEL---}} #1
}

\title{Scale-robust Auctions\thanks{
An extended abstract of this work has appeared in 61st Annual Symposium on Foundations of Computer Science (FOCS'20) under the title ``Benchmark Design and Prior-independent Optimization". 
The authors thank Tan Gan, Shengwu Li, Eran Shmaya and Philipp Strack for helpful comments and suggestions. 
Lemma 6 of an earlier version of this paper, which claimed that it is without loss to restrict attention to mechanisms that never allocate to the lower valued bidder, was incorrect.  This error was pointed out by \citet{anunrojwong2026robust} who also exhibits a mechanism with strictly better performance than ours that sometimes allocates to the lowest valued bidder.  In the present version of this paper, the error is corrected and our main result establishes robust optimality only within the class of mechanisms that does not allocate to the lower valued bidder.
}}

\newcommand{\email}[1]{\href{mailto:#1}{#1}}

\author{
Jason Hartline\thanks{Northwestern University. Email: \email{hartline@northwestern.edu}. } \and
Aleck Johnsen\thanks{Northwestern University.  Email: \email{aleckjohnsen@u.northwestern.edu}. } \and 
Yingkai Li\thanks{National University of Singapore. Email: \email{yk.li@nus.edu.sg}. }
}
\date{}

\maketitle
\begin{abstract}
We study auctions that are robust at any scale, i.e., they can be applied to sell both expensive and cheap items and achieve the best multiplicative approximation of the optimal revenue in the worst case.
We first show that it is without loss of optimality to restrict attention to scale-invariant mechanisms whenever the family of possible distributions is closed under every positive rescaling.
This conclusion uses no regularity or other distributional shape restriction.
We then solve the two-agent, single-item problem with values drawn i.i.d.\ from an unknown regular distribution when only a high value bidder can receive a positive allocation. The robustly optimal mechanism in this class randomizes between the second-price auction, with probability approximately $\asimp$, and a markup auction that offers the item to the highest-valued bidder at a price equal to $\rsimp$ times the second-highest value. Its worst-case approximation ratio is approximately $\apxsimp$.
\end{abstract}


\JEL{D44, D82}

\section{Introduction}
\label{s:intro}
In many markets, it is customary to implement fixed proportional transaction fees regardless of the scale of the commodity. 
For instance, in real estate, agents typically charge a commission fee of around 6\% regardless of the sale price of the house. 
Similarly, in digital application markets, the Apple Store imposes a 30\% service fee, while Google Play charges a 15\% service fee for each app purchase, irrespective of the app's price. 
Motivated by this feature of markets, we consider the design of auctions that are resilient to scale, i.e., that achieve a favorable revenue guarantee approximating the optimal revenue in a multiplicative manner.

We study a robust analysis framework in which the principal designs auctions that perform well at all scales \citep{hartline2008optimal}. In this framework, the principal seeks an auction that is independent of the distribution over agents' values and, specifically, the scale of the distribution. 
The goal is to minimize the multiplicative approximation of the optimal mechanism in the worst case over a family of possible distributions.\footnote{This analysis framework is known as prior-independent approximation in the computer science literature following \citet{hartline2008optimal}}

We study the single-item auction in a symmetric environment 
where the buyers' values are drawn independently and identically from a regular distribution.\footnote{\label{foot:regularinformal}A distribution is regular if its corresponding virtual value function is non-decreasing \citep{mye-81}.}  
For regular distributions,
if the distribution is known by the principal, 
the second price auction with monopoly reserve is Bayesian optimal \citep{mye-81}.
If the regular distribution is unknown, \citet{BK-96} show that by adding an additional buyer, the seller can extract at least the optimal revenue (without the additional buyer) using the second-price auction.
A corollary of this result is that with a fixed market size of $n$ buyers, 
the second-price auction attains at least $1-\frac{1}{n}$ fraction of the optimal revenue. 
Thus, in large markets where the number of buyers converges to infinity, the second-price auction is asymptotically optimal, 
while in small markets, the multiplicative gap between the optimal revenue and the second-price auction can be as bad as~2. 
Is the second-price auction, via this corollary of Bulow and Klemperer, the best scale-robust auction?

In this paper, we focus on the design of optimal scale-robust mechanisms in small markets.  In particular, we focus on the extreme case proposed by \citet{DRY-15} where there are only two buyers.  
The restriction to small markets is consistent with our motivation of robust analysis. 
Unlike in large markets, where sellers can rely on abundant historical data to accurately estimate the valuation distributions of buyers, 
such data is insufficient in small markets.
Therefore, a seller with limited information often finds it natural to adopt the scale-robust approach for selling the goods. 
When there are only two buyers, \citet{AB-18} show that the second-price auction is indeed scale-robust optimal if the valuation distribution of the buyers satisfies the monotone hazard rate condition (MHR). 
However, \citet{FILS-15} show that the seller can improve her worst case approximation guarantee by randomly marking up the second-price 
if the valuation distribution only satisfies the regularity condition 
(which is weaker than MHR). 
The main intuition is that without MHR, the worst-case valuation distribution may be too heavy-tailed, 
and hence the seller benefits from randomization to 
hedge between the case in which the second-price auction is optimal 
and the case in which the monopoly price is much higher than the second-price. 

We call a mechanism \emph{highest-value-only} if it assigns positive allocation probability only to a highest-valued buyer; the item may be withheld. We identify the robustly optimal dominant-strategy incentive-compatible (DSIC) mechanism in this class for two buyers with values drawn i.i.d.\ from a regular distribution.
The optimal highest-value-only mechanism is a mixture of the second-price auction
and a markup auction that offers the highest-valued buyer the item at a price equal to
approximately 2.45 times the other buyer's value.\footnote{An alternative view of the optimal highest-value-only mechanism is that the winning agent only receives the full item if his bid is sufficiently high compared to the second-highest bid,
and receives a ``damaged'' item, or equivalently a partial allocation of the item,
if his bid is close to the second-highest bid.}
The second-price auction is used with probability approximately $\asimp$ and the markup auction with the remaining probability, and the resulting worst-case approximation ratio is approximately $\apxsimp$.
This pins down the optimal ratio within the highest-value-only class, up to the numerical precision documented in \Cref{app:prior-independent}. Follow-up work by \citet{anunrojwong2026robust} constructs a DSIC mechanism that sometimes allocates to a lower-valued buyer and provides a strictly better guarantee, so the highest-value-only restriction is costly; the exact unrestricted optimum remains open.

We nevertheless focus on highest-value-only mechanisms because they define a natural and practically important constrained frontier.
Their defining principle is simple: whenever the item is allocated, only a highest-valued buyer may receive it. They therefore respect value ranking even when they randomize or sometimes withhold the item.
Analogous highest-bidder rules appear in practice. In eBay auction-style listings, the highest bidder wins whenever the item is sold, while automatic bidding raises a bidder's offer only as needed to remain ahead
of competing bids \citep{roth2002last}. 
In U.S.~offshore oil and gas lease sales, BOEM evaluates high bids under a bid-adequacy process designed to ensure that the government receives fair market value for each lease issued \citep{boemBidAdequacy2025}.
These institutions are not literal implementations of our model, but they illustrate the appeal of
an allocation rule that may withhold the item while never selecting a lower bid over a higher one.
Finally, the highest-value-only class includes the second-price and relative-markup
auctions at the center of prior-independent design, and in our setting its optimum is attained by a transparent mixture of these familiar formats. 
Characterizing this frontier therefore isolates what can be achieved without allocating to a lower-valued buyer and provides a sharp benchmark for measuring the revenue value of allowing such allocations.

Note that our restriction to DSIC mechanisms is not without loss of generality when the seller can adopt more general and potentially non-truthful mechanisms \citep[e.g.,][]{caillaud2005implementation,feng2021revelation}. 
However, we aim to design auctions that are robust to the beliefs of all parties, and DSIC mechanisms provide max-min optimal revenue guarantees over the worst-case beliefs of the buyers \citep{chung2007foundations}. 

Our characterization has two main takeaways.
First, scale invariance can be imposed without loss.
This is based on the observation that a crucial uncertainty we guard against is a common multiplicative rescaling of values (i.e., a change of units or an inflation shock). Any mechanism that embeds fixed dollar thresholds, entry fees, or caps is fragile: by choosing the units, the environment can push the mechanism into its weakest regime. Scale-invariant designs remove this lever: allocations depend only on relative magnitudes or order statistics, and transfers scale proportionally with values. 
Practically, the same mechanism is then portable across markets and currencies without retuning any threshold.
\Cref{sec:scale_invariant} shows, under mild conditions, that the restriction to scale-invariant mechanisms is without loss whenever the family of possible distributions contains every positive rescaling of each of its members.
The argument uses no regularity or other distributional shape restriction. Our result also does not require the allocation rule to have a limit as all reports are proportionally scaled toward zero; the reductions in \citet{AB-18} and in the follow-up work by \citet{anunrojwong2026robust} rely on such a limit.

Second, we show that optimal robust performance within the highest-value-only class requires randomization between the second-price auction and a single markup price, rather than a continuum of markup prices.
This result stems from the balance between two opposing forces. When the markup strictly exceeds 1, no matter how small, the seller risks losing sales, leading to a discontinuous drop in revenue compared to a mechanism without markups. To justify this risk, the markup must be sufficiently high to generate substantial revenue when buyers' valuations are sufficiently dispersed. However, if the markup is too large, the probability that buyers are sufficiently apart diminishes, making the expected benefit again insufficient to offset the revenue loss from foregone trades.
Our results identify that there exists a unique intermediate markup price that optimally balances these effects, ensuring robust performance. 

\subsection{Related Work}
\label{sec:related-work}

The scale-robust framework provides a natural worst-case criterion for comparing prior-independent mechanisms. Prior work characterized optimal mechanisms in several restricted environments but did not identify the robustly optimal highest-value-only mechanism for two bidders with i.i.d.\ regular values.
\citet{HR-14} provided the optimal mechanism for revenue
maximization in the sale of a single item to a single agent with value
from a bounded support, where the optimal mechanism
posts a randomized price. 
For revenue maximization in the sale of an item to
one of two agents with values drawn from an i.i.d.\ regular distribution, \citet{DRY-15} show that the second price auction
is a 2-approximation.  \citet{FILS-15} provided a randomized mechanism showing that 
this factor of~2 is not tight. 
Upper and lower bounds on this canonical
problem were improved by \citet{AB-18} to be within $[1.80,1.95]$ for DSIC mechanisms. 
The main result of our paper identifies the robustly optimal highest-value-only mechanism for this environment, with an approximation factor of about $1.91$.
The optimal randomization has a particularly simple form: it mixes the second-price auction with a single markup price strictly above 1.
Finally, for this  two-agent problem with i.i.d.\ values from a
distribution in the subset of regular distributions that further satisfy a monotone hazard rate condition, \citet{AB-18} show that the
second-price auction is scale-robust optimal.  

In follow-up work, \citet{anunrojwong2026robust} studies the same
scale-robust objective and builds on the stochastic-markup
characterization developed in this paper. Expressed as a fraction of the
distribution-specific optimal revenue, our highest-value-only mechanism
guarantees approximately $0.524413$. The follow-up paper then
constructs a ``random-favorite'' mechanism that sometimes
allocates to the lower-valued buyer and guarantees approximately
$0.524829$. Thus imposing the highest-value-only restriction is strictly costly under regularity,
although the known quantitative improvement is small. The exact
unrestricted optimum remains unresolved.

The restriction to DSIC mechanisms has the desirable property that agents' behaviors and the expected revenue in DSIC mechanisms 
do not rely on agents' information about each other,
and the set of DSIC mechanisms is equivalent to the set of ex post implementable mechanisms \citep{bergemann2005robust}. 
Without the restriction to DSIC mechanisms, \citet{caillaud2005implementation} 
use an ascending auction in virtual value space to implement the Bayesian optimal mechanism. 
A critique of such implementation is that this mechanism takes the common knowledge assumption too literally
and is impractical for real-world applications. 
\citet{FH-18} and \citet{feng2021revelation} show that there exist simple and practical non-incentive-compatible mechanisms that outperform the optimal DSIC mechanism, 
and further study of non-incentive-compatible mechanisms is still warranted within the scale-robust analysis framework.

Our paper relates to the auction design literature with max-min optimal and min-max regret objectives
when the principal is ignorant of the value distribution.
For max-min optimization, \citet{bergemann2011robust} and \citet{carrasco2018optimal}
consider the design of a robustly optimal mechanism in the single-item, single-buyer setting. 
\citet{NI22} extend the model to two i.i.d.\ buyers 
and the model with correlated valuations is considered in \citet{che2022robustly}.
Both papers identify the second-price auction with random markups as the max-min optimal mechanism, 
where the distribution over markups relies on the expected value of each buyer.
By contrast, the information about the expected value is not available to the principal in our model, 
and there exists a fixed distribution over markups that achieves the optimal approximation ratio.

For min-max regret optimization, 
the optimal distribution over prices for the single-item, single-buyer setting is characterized in \citet{bergemann2008pricing,bergemann2011robust}.
\citet{anunrojwong2022robustness} show that a second-price auction
with random reserve prices is robustly optimal when there are multiple agents, even if the values of the agents can be correlated. 

Robust approaches to mechanism design and pricing have received considerable attention in the operations literature, where a policy must be committed to before the environment is known.
This includes distributionally robust mechanism design under support and moment information \citep[e.g.,][]{koccyiugit2020distributionally}, robust monopoly pricing with limited demand information \citep[e.g.,][]{eren2010monopoly}, and prior-independent auctions and pricing from samples \citep[e.g.,][]{AB-18,allouah2022pricing}.

\section{Model}
\label{sec:prelim}
The principal sells a single item to $n=2$ agents with private values $\vals =
(\vali[1],\vali[2])$. 
The agents have linear utilities,
i.e., agent $i$'s utility is $\vali\,\alloci - \pricei$ for allocation
probability $\alloci$ and expected payment $\pricei$.  Agents' values
are drawn independently and identically from a product distribution
$\dists = \dist \times \dist$ where $\dist$ will denote
the cumulative distribution function of each agent's value.

\paragraph{Mechanisms} 
A mechanism $\mecha$ is defined by an ex post allocation and payment
rule $\allocs^{\mecha}$ and $\prices^{\mecha}$ which map the profile
of values $\vals$ to a profile of allocation probabilities and a
profile of payments, respectively.  We focus on mechanisms that are
feasible, dominant strategy incentive compatible, and individually
rational:
\begin{itemize}
  \item For selling a single item, a mechanism is {\em feasible} if for all
    valuation profiles, the allocation probabilities sum to at most
    one, i.e., $\forall \vals,\ \sum_i\alloci^{\mecha}(\vals) \leq 1$.
  \item A mechanism is {\em dominant strategy incentive compatible} if
    no agent $i$ with value $\vali$ prefers to misreport some value
    $z$: $\forall \vals,i,z,\ \vali\,\alloci^{\mecha}(\vals) -
    \pricei^{\mecha}(\vals) \geq \vali\,\alloci^{\mecha}(z,\valsmi) -
    \pricei^{\mecha}(z,\valsmi)$ where $(z,\valsmi)$ denotes the
    valuation profile with $\vali$ replaced with $z$.
  \item A mechanism is {\em individually rational} if truthful
    reporting always leads to non-negative utility: $\forall
    \vals,i,\ \vali\,\alloci^{\mecha}(\vals) - \pricei^{\mecha}(\vals)
    \geq 0$.
\end{itemize}

\begin{definition}[Highest-Value-Only Mechanism]\label{def:highest-value-only}
A two-agent auction is \emph{highest-value-only} if it assigns positive
allocation probability only to a highest-valued agent. Equivalently,
for every valuation profile $\vals=(\vali[1],\vali[2])$ and each
$i\neq j$, $\vali<\vali[j]$ implies
$\alloci^{\mecha}(\vals)=0$.
\end{definition}
The highest-value-only restriction is an assumption on the designer's
choice set, not a consequence of feasibility, incentive compatibility,
individual rationality, or scale robustness.

Denote a family of feasible mechanisms by
$\feasibleMecha$ and a mechanism in this family by $\mecha$. 
The expected revenue of mechanism $\mecha$ when the value profile is $\vals$
is denoted by $\mecha(\vals)$. 
When evaluating the revenue of a mechanism
in expectation over the distribution,
we adopt the short-hand notation 
$\mecha(\dist) = \expect[\vals \sim \dist\times\dist]{\mecha(\vals)}$.
Given a distribution $\dist$ and a family of mechanisms $\feasibleMecha$, 
we define the optimal-revenue benchmark as the value
\begin{align*}
\OPT_{\dist}(\dist) := \sup_{\mecha \in \feasibleMecha}\mecha(\dist).
\end{align*}
When the supremum is attained, $\OPT_{\dist}$ may also denote an
optimizing mechanism. Defining the benchmark by a supremum avoids an
unnecessary attainment assumption for the nonregular distribution
families allowed in \Cref{thm:scale_invariant_is_opt}.

\paragraph{Revenue Curves} A mechanism's revenue can be easily and geometrically understood via the
marginal revenue approach of \citet{mye-81} and \citet{BR-89}.
For distribution $\dist$, the \emph{quantile} $\quant$ of an agent
with \ value $\val$ denotes how strong that agent is relative to
the distribution $\dist$. 
Specifically, quantiles are defined by the mapping
$\quantf_{\dist}(\val) = \Pr_{z\sim\dist}\{z\geq \val\}$. 
Denote the mapping back to
value space by $\valf_{\dist}$, i.e., $\valf_{\dist}(\quant)$ is the
value of the agent with quantile~$\quant$. 
A single agent {\em price-posting revenue curve} $\pRev(\quant)$ gives the revenue of posting a price 
such that the probability that the agent accepts the price is $\quant$. 
For an agent with value distribution $\dist$, price $\valf_{\dist}(\quant)$ is accepted with probability $\quant$,
and its expected revenue is $\pRev(\quant) = \quant\cdot\valf_{\dist}(\quant)$. 
A single agent {\em revenue curve} $\rev_{\dist}(\quant)$ gives the optimal revenue from selling to a single agent
under the constraint that ex ante sale probability is $\quant$. 
By \citet{BR-89}, the revenue curve $\rev$ is always concave,
and it coincides with the concave hull of the price-posting revenue curve~$\pRev$. 
In this paper, we focus on the family of regular distributions. 
\ 
Let $\regular$ be the family of i.i.d.\ regular value distributions with finite first moment.
\begin{assumption}[Regularity]
A distribution $\dist$ is regular if the price-posting revenue curve~$\pRev$ is concave.\footnote{An equivalent definition for regularity is that the virtual value function 
$\virtual(\val) = \val - \frac{1-\dist(\val)}{\density(\val)}$
is non-decreasing in $\val$.}
\end{assumption}

An immediate implication \ for a regular distribution is that 
the price-posting revenue curve coincides with the revenue curve, i.e., $\pRev=\rev$.
The optimal mechanism for a single agent
posts the {\em monopoly price} $\valf_{\dist}(\qo)$ which corresponds to
the monopoly quantile $\qo = \argmax_\quant \rev_{\dist}(\quant)$.
In multi-agent settings, the expected revenue of any multi-agent mechanism $\mech$ equals
its expected surplus of marginal revenue.

\begin{lemma}[\citealp{mye-81}]
  \label{thm:myerson}
  Given any incentive-compatible mechanism $\mecha$ 
with allocation rule $\allocs^{\mecha}(\vals)$, 
the expected revenue of mechanism $\mecha$
for agents with regular distribution $\dists$ is 
equal to its expected surplus of marginal revenue, 
i.e., 
\begin{equation*}
  \mecha(\dists) = \sum\nolimits_i \expect[\vals \sim \dists]{\pricei^{\mecha}(\vals)} = \sum\nolimits_i \expect[\vals \sim \dists]{
\rev'_{\dist}(\quantf_{\dist}(\vali))\cdot\alloci^{\mecha}(\vals) 
}.
\end{equation*}
\end{lemma}

\begin{corollary}[\citealp{mye-81}]
For i.i.d., regular, single-item environments, the optimal mechanism
$\OPT_{\dist}$ is the second-price auction with anonymous reserve equal to the
monopoly price.
\end{corollary}

\paragraph{Robust Objectives} In this paper, we consider the model where the principal is ignorant of the true distribution over values. 
Instead, the principal knows that the true distribution belongs to a family $\feasibleDist$ 
and designs a mechanism that minimizes the worst case approximation ratio to the optimal revenue for distributions within $\feasibleDist$. 
This scale-robust analysis framework is also referred to as the prior-independent mechanism design \citep{hartline2008optimal}.

\begin{definition}[Robust Framework]
\label{def:pimd}
  The {\em scale-robust analysis framework} is given by a
  family of mechanisms $\feasibleMecha$ and a family of distributions
  $\feasibleDist$ and solves the program
\begin{align}
  \label{eq:pi}
  \tag{$\piratio$}
\piratio &\triangleq \;\inf_{\mecha \in \feasibleMecha} \sup_{\dist \in \feasibleDist}
\frac{\OPT_{\dist}(\dist)}{\mecha(\dist)}.
\end{align}
\end{definition}
In our paper, we focus on the family of mechanisms $\feasibleMecha$ that are dominant strategy incentive compatible and individually rational. 
The scale-invariance result in \Cref{thm:scale_invariant_is_opt} imposes measurable allocation and payment rules and a finite first moment for every admissible value distribution, which makes payments bounded by values admit an integrable envelope. These technical conditions are not regularity or other distributional shape restrictions. For the main mechanism-characterization result, $\regular$ incorporates the finite-moment condition.

\section{Optimality of Scale Invariance}
\label{sec:scale_invariant}
We first show that for the scale-robust analysis framework, it is without loss of optimality to focus on robust mechanisms that are scale invariant. 

\begin{definition}[Scale-closed Distribution Family]\label{def:scale-closed}
A family $\feasibleDist$ of one-dimensional value distributions is
\emph{scale closed} if, for every $\dist\in\feasibleDist$ and every
$s>0$, the law $\dist^{s}$ of $sV$ for $V\sim\dist$ also belongs to
$\feasibleDist$.
\end{definition}

\begin{definition}[Scale Invariant]
  Given any incentive-compatible mechanism $\mecha$ with allocation rule
  $\alloc^{\mecha}(\vals)$, mechanism $\mecha$ is \emph{scale invariant}
  if for each agent $i$, valuation profile $\vals$ and any constant
  $\valscale > 0$, $\alloci^{\mecha}(\valscale \cdot \vals) =
  \alloci^{\mecha}(\vals)$.   For a normalized DSIC payment rule, meaning $\pricei^{\mecha}(0,\valsmi)=0$, the payment identity implies $\pricei^{\mecha}(a\vals)=a\pricei^{\mecha}(\vals)$ and hence $\mecha(a\vals)=a\mecha(\vals)$.
\end{definition}

\begin{theorem}\label{thm:scale_invariant_is_opt}
Let $\feasibleDist$ be any scale-closed family of value
distributions with finite first moment, and take
allocation and payment rules to be Borel measurable in this two-agent environment. If, for some $\rho\geq0$, a feasible, DSIC, and ex-post IR mechanism $\mecha$
satisfies
\[
  \mecha(\dist)\geq \rho\,\OPT_{\dist}(\dist)
  \qquad\text{for every }\dist\in\feasibleDist,
\]
then there exists a feasible, DSIC, ex-post IR, and scale-invariant
mechanism $\widehat{\mecha}$ satisfying the same inequalities. Thus
$\widehat{\mecha}$ has an approximation ratio no worse than that of
$\mecha$. If $\mecha$ is highest-value-only, $\widehat{\mecha}$ can also be chosen highest-value-only.
\end{theorem}
In our scale robust framework, we judge a mechanism by its worst-case performance over all multiplicative rescalings of agents' values. 
Under this criterion, scale-invariant designs are the natural fixed points. 
Two complementary intuitions make this compelling.

\smallskip\noindent\emph{(1) Units-of-measure neutrality and immunity to an adversarial scale.}
If the designer's guarantee can be improved or worsened by expressing values in dollars rather than cents (or by an inflation shock that multiplies all valuations), then the guarantee is partly an artifact of units. 
Worse, in a max--min evaluation nature effectively chooses the units, so any fixed dollar threshold, entry fee, or cap is a lever the adversary can pull to place the mechanism in its least favorable regime.
Scale-invariant mechanisms eliminate this artifact: allocations depend only on \emph{relative} magnitudes, and payments scale proportionally. 
In the same way that ad valorem taxes are neutral to the price level whereas specific (per-unit) taxes are not, scale-invariant mechanisms are neutral to the choice of units, and thus their guarantees reflect economics rather than accounting.

\smallskip\noindent\emph{(2) Symmetry suggests symmetrization.}
The uncertainty we are guarding against---a common multiplicative shock to all values---is a symmetry of the environment: it changes the units but not the economically meaningful comparisons among types. 
In robust design, a standard heuristic is that optima respect the symmetries of the uncertainty set. 
Intuitively, if a mechanism reacts differently across scales, we can ``wash out'' that sensitivity by averaging its behavior across scales; this removes avoidable variability in performance without sacrificing its best features. 
What remains is a scale-invariant mechanism that is at least as safe against scale misspecification.


\paragraph{Proof Sketch.}
For $t\in\mathbb R$, scale the reports by $e^t$ and divide payments by
$e^t$. Scale closure is exactly what lets the guarantee for $\mecha$
be applied to the scaled distribution. Averaging these mechanisms over
$t\in[-m,m]$ therefore preserves the guarantee and is asymptotically
scale invariant. Anonymity and approximate scale invariance reduce the allocation rule
to a one-dimensional monotone function of the bid ratio. Helly's
selection argument then gives a pointwise-convergent subsequence. The
normalized payment identity gives pointwise convergence of payments,
and the bound
$0\leq p_i(\vals)\leq \vali$ supplies the integrable envelope for the
Dominated Convergence Theorem. The resulting limit is exactly scale
invariant and retains the same guarantee. See
\Cref{apx:scale_invariant} for details.

\section{Optimal Scale-Invariant Mechanisms under the Highest-Value-Only Restriction}
\label{sec:prior-independent}

We solve the scale-robust revenue problem under the following restrictions:
\begin{itemize}
\item the family of i.i.d.\ regular value distributions $\regular$; and
\item 
  the family of feasible, incentive-compatible, individually
  rational, scale-invariant, and highest-value-only mechanisms.
\end{itemize}

\begin{assumption}[Highest-Value-Only Mechanisms]\label{ass:highest-value-only}
For the mechanism-characterization result, the designer's choice set is
restricted to highest-value-only mechanisms in the sense of
\Cref{def:highest-value-only}: when the reported values differ, only the
higher-valued agent may receive the item.
\end{assumption}

The following discussion motivates these restrictions in the robust design problem.  
\begin{itemize}
\item \emph{Restrictions on Distributions.} 
In general, without assumptions of symmetric or regular value distributions, no mechanism achieves good scale-robust performance. Analyzing robust mechanism design without these assumptions does not allow us to distinguish good mechanisms from bad ones, as all mechanisms perform poorly in such settings. 
Moreover, nearly all papers on the scale-robust analysis framework focus on i.i.d.\ agents, and almost all studies on revenue maximization within this framework restrict attention to regular distributions. 

\item \emph{Restrictions on Mechanisms.}
The restriction to feasible and individually rational mechanisms is necessary to have a sensible optimization problem. The restriction to incentive compatibility is standard in almost all papers on the scale-robust analysis framework, with the exception of \citet{FH-18}, which shows that this restriction is not without loss of optimality. However, maintaining incentive compatibility ensures robustness to agents' beliefs as well. Additionally, the assumption of scale invariance is without loss (\cref{thm:scale_invariant_is_opt}), which simplifies the structure of the robust mechanisms. By contrast, \Cref{ass:highest-value-only} is a substantive restriction: follow-up work by \citet{anunrojwong2026robust} shows that it strictly lowers the best robust guarantee under regularity.
\end{itemize}

In our environment, stochastic markup mechanisms form a scale-invariant subclass of the lookahead mechanisms of \citet{ron-01}.

\begin{definition}[Markup Mechanism]
\label{d:markup}
For any parameter $r\geq 1$, the {\em $\ratio$-markup mechanism} $\mecha_{\ratio}$ identifies the
agent with the highest value (with ties broken uniformly at random)
and offers this agent $\ratio$ times the second-highest value. A
{\em stochastic markup mechanism} draws~$\ratio$ from a given
distribution on $[1,\infty)$.  The family of stochastic markup mechanisms is $\stocmarkup$.
\end{definition}

Notice that the second-price auction is the $1$-markup mechanism~$\mecha_1$, so the family $\stocmarkup$ contains the second-price auction as a special case.
Our main result will show that the robustly optimal highest-value-only mechanism is a stochastic markup mechanism, and we identify the optimal distribution of markups for regular distributions.

In the analysis, given our restriction
to scale-invariant mechanisms, it will be sufficient to consider
distributions that are normalized so that the single-agent optimal
revenue is $\max_\quant \rev(\quant) = 1$.
An important family of distributions with revenue normalized to 1 is the normalized triangle distributions, which have
revenue curves shaped like triangles (\Cref{f:quad}). 
For the stochastic markup mechanisms relevant to our characterization, these distributions capture the worst-case regular distributions.

\begin{figure}[t]
\begin{flushleft}
\begin{minipage}[t]{0.48\textwidth}
\centering
\newcommand{\TRISCALE}{0.7}

\begin{tikzpicture}[scale = \TRISCALE]

\draw (-0.2,0) -- (10, 0);
\draw (0, -0.2) -- (0, 5.5);

\draw (0, 0) -- (3, 5);
\draw (3, 5) -- (9, 0);

\draw [dotted] (0, 5) -- (3, 5);
\draw [dotted] (3, 0) -- (3, 5);

\draw (-0.5, 5) node {$1$};

\draw (0, -0.5) node {$0$};
\draw (9, -0.5) node {$1$};
\draw (3, -0.5) node {$\qo$};

\end{tikzpicture}
\end{minipage}
\begin{minipage}[t]{0.48\textwidth}
\centering
\newcommand{\QUADSCALE}{0.7}

\begin{tikzpicture}[scale = \QUADSCALE]

\draw (-0.2,0) -- (10, 0);
\draw (0, -0.2) -- (0, 5.5);

\draw (0, 0) -- (3, 5);
\draw (3, 5) -- (6, 4);
\draw (6, 4) -- (9, 0);

\draw [dotted] (0, 5) -- (3, 5);
\draw [dashed] (0, 0) -- (7.2, 4.8);
\draw [dotted] (3, 0) -- (3, 5);

\draw [dotted] (6, 0) -- (6, 4);

\draw (-0.5, 5) node {$1$};

\draw (0, -0.5) node {$0$};
\draw (9, -0.5) node {$1$};
\draw (3, -0.5) node {$\qo$};
\draw (6, -0.5) node {$\qt$};

\draw (6.5, 5) node {$\sfrac{1}{r\qo}$};

\end{tikzpicture}
\end{minipage}
\vspace{-18pt}
\end{flushleft}
\caption{\label{f:quad}
The \ left-hand side is the revenue curve for triangle distribution $\tri_{\qo}$
and the \ right-hand side is the revenue curve for quadrilateral distribution $\Qr_{\qo, \qt, \ratio}$. 
The definition of quadrilateral distribution $\Qr_{\qo, \qt, \ratio}$ will be formally introduced later in \Cref{sec:best response}.
}
\end{figure}

\begin{definition}[Triangle Distribution]\label{def:triangle}
A \emph{normalized triangle distribution with monopoly quantile $\qo$},
denoted $\tri_{\qo}$, is defined by the quantile function
\begin{equation*}
\quantf_{\tri_{\qo}}(\val) =
\begin{cases}
\frac{1}{1 + \val(1-\qo)} & \val \leq
\sfrac{1}{\qo}\\ 0 & \text{otherwise.}
\end{cases}
\end{equation*} The {\em
    triangulation} of a normalized distribution with monopoly quantile
$\qo$ is $\tri_{\qo}$.  The family of normalized triangle
distributions with finite first moment is $\TRIF = \{\tri_{\qo} : \qo \in (0,1]\}$. We also write $\tri_0$ for the equal-revenue boundary limit. It has infinite first moment and is therefore not an admissible member of $\regular$; it is used below only to evaluate limits at the boundary of $\TRIF$.
\end{definition}
Intuitively, for any monopoly quantile $\qo$, 
\ the normalized triangle distribution is the distribution that is first order stochastically dominated by any other distribution with monopoly quantile $\qo$. 
That is, in the single-agent problem, 
the normalized triangle distribution minimizes the expected revenue of any given mechanism while maintaining the optimal revenue and monopoly quantile unchanged. 

\begin{theorem}\label{thm:pi optimal mechanism}
Under \Cref{ass:highest-value-only}, for i.i.d., regular, two-agent,
single-item environments, the robustly optimal mechanism among
highest-value-only mechanisms that are feasible, DSIC, and ex-post IR is
$\mechaARO$,
which randomizes over the second-price auction $\mecha_1$ with
probability $\optweight$ and $\optratio$-markup mechanism
$\mecha_{\optratio}$ with probability $1-\optweight$, where
$\optweight \approx \asimp$ and $\optratio\approx \rsimp$. The
worst-case regular distribution for this mechanism is triangle distribution
$\tri_{\optqo}$ with $\optqo \approx \qsimp$ and its approximation
ratio is $\beta \approx \apxsimp$.
\end{theorem}

The parameters of the optimal highest-value-only mechanism solve a
transcendental system and are therefore reported numerically.
\Cref{app:prior-independent} certifies optimality within the
highest-value-only class up to an additive tolerance of $2\times 10^{-8}$ in
the approximation ratio; this tolerance comes entirely from numerical
evaluation.

In the two sections below, we prove this theorem with the following
main steps. 
\begin{enumerate}
\item 
We first analyze the auxiliary markup--triangle game in which
the designer selects a stochastic markup mechanism and nature selects
a triangle distribution. \Cref{thm:triangle} characterizes its
equilibrium.

\item 
\Cref{lem:perturbed-reduction} transfers the resulting lower
bound to every scale-invariant mechanism satisfying the highest-value-only restriction by weakly dominating
its revenue on atomless i.i.d.\ distributions with a stochastic markup
mechanism and using atomless regular approximations to the equilibrium
triangle. Conversely,
\Cref{thm:tri} shows that triangle distributions are worst case among
regular distributions for the relevant stochastic markup mechanisms.
\end{enumerate}
\Cref{thm:scale_invariant_is_opt} and anonymization first permit
attention to be confined to symmetric, scale-invariant, highest-value-only
mechanisms. The two steps above then give the matching lower and upper
bounds and prove \Cref{thm:pi optimal mechanism}.

\subsection{Stochastic Markup Mechanisms versus Triangle Distributions}
\label{sub:triangle}

In this section, we analyze the auxiliary markup--triangle game in
which the designer chooses a stochastic markup mechanism and nature
chooses a triangle distribution.
We first define a general family of truncated distributions, which will be
important subsequently in the proof. 
Recall that for scale-invariant mechanisms, 
it is without loss of generality to normalize the distributions to have monopoly revenue of one.

\begin{definition}[Truncated Distribution]
  \label{def:truncate}
  A distribution is {\em truncated} if the highest-point in its
  support is the monopoly price (typically a point mass).  The {\em
    truncation} of a distribution is the distribution that replaces
  every point above the monopoly price with the monopoly price.  The
  family of truncated distributions is denoted $\TRUNCF$.
\end{definition}

\paragraph{Revenue of Various Mechanisms} 
We will provide three lemmas below to present the formulae for the revenue of
the optimal mechanism, the second-price auction, and non-trivial
markup mechanisms for triangle distributions.  The formula for
the revenue of markup mechanisms is discontinuous at $\ratio = 1$.  Thus,
in our discussion, we will distinguish between the second-price auction
$\mech_1$ and the non-trivial markup mechanism $\mech_\ratio$ for $\ratio
> 1$.

Before the details of those formulations, we would like to first introduce a technical lemma from \citet{DRY-15}, which follows immediately from
\Cref{thm:myerson} and provides a geometric understanding of the expected revenues of the second-price auction and the optimal mechanism
in two-agent settings. The geometry is illustrated in \cref{f:illustrate_rev}. 
Intuitively, the revenue from each agent can be interpreted as being generated by a mechanism determined by the reported values of the other agents. For example, in a second-price auction, each agent effectively faces a distribution over posted prices, where the price always equals the reported value of the other agent. This perspective is particularly useful for our analysis.
\begin{figure}[t]
\centering
\newcommand{\REVONESCALE}{0.6}

\begin{tikzpicture}[scale = \REVONESCALE]

\draw [name path global = A] plot [smooth, tension=0.45] coordinates {(0, 0) (0.5, 2.5) (1,3.5) (1.9, 4.4) (2.5, 4.8) (3, 5)};

\draw [name path global = B, dotted] (3, 0) -- (3, 5);

\tikzfillbetween[of=A and B]{gray!40!white};

\fill[color=gray!40!white] (3, 0.01) -- (3, 5) -- (9, 5) -- (9, 0.01);

\draw [thick] plot [smooth, tension=0.5] coordinates {(3, 5) (4.5, 4.76) (6, 4) (8, 2) (9, 0)};
\draw [thick] plot [smooth, tension=0.45] coordinates {(0, 0) (0.5, 2.5) (1,3.5) (1.9, 4.4) (2.5, 4.8) (3, 5)};
\draw [dotted] (0, 5) -- (3, 5);
\draw [dotted] (3, 0) -- (3, 5);

\draw (-0.9, 5) node {$\rev(\qo)$};

\draw (0, -0.5) node {$0$};
\draw (9, -0.5) node {$1$};
\draw (3, -0.5) node {$\qo$};

\draw (-0.1,0) -- (10, 0);
\draw (0, -0.1) -- (0, 5.5);

\end{tikzpicture}
\caption{\label{f:illustrate_rev}
The solid black curve is the revenue curve $\rev(\quant)$ for the single-agent setting. 
The gray area is the area under the smallest monotone concave upper bound of the revenue curve, 
which is half of the optimal revenue. 
}
\end{figure}

\begin{lemma}[\citealp{DRY-15}]
\label{l:DRY-15}
In i.i.d.\ two-agent single-item environments, 
\begin{itemize}
\item the expected revenue of second-price auction 
is twice the area under the revenue curve;

\item the expected revenue of the optimal mechanism is twice the
area under the smallest monotone concave upper bound of the revenue curve.
\end{itemize}
\end{lemma}

Next, we present the three lemmas for the revenues of various mechanisms. 

\begin{lemma}\label{lem:opt rev}
  For i.i.d., normalized truncated, two-agent, single-item
  environments, the optimal mechanism posts the monopoly price and
  obtains revenue $2-\qo$, where $\qo$ is the probability that an
  agent's value equals the monopoly price.
\end{lemma}

\begin{proof}
  The smallest monotone concave function that upper bounds the revenue
  curve is a trapezoid; its area is $\sfrac{\qo}{2} + 1-\qo$.  The
  optimal revenue from two agents, by \Cref{l:DRY-15}, is twice this
  area, i.e., $2-\qo$.\footnote{Equivalently, it is also easy to verify that for truncated distributions, the optimal mechanism is to post a price of $\frac{1}{\qo}$ to both agents, which yields an expected revenue of $2-\qo$. }
\end{proof}

\begin{lemma}
  \label{lem:spa rev}
The revenue of the second-price auction $\mech_1$ for distribution
$\tri_{\qo}$ is 1, i.e., $\mech_1(\tri_{\qo}) = 1$.
\end{lemma}

\begin{proof}
  By \Cref{l:DRY-15}, the revenue is twice the area under the revenue
  curve.  That area is $\sfrac 1 2$; thus, the revenue is 1.
\end{proof}

\begin{lemma}\label{clm:rev of r}
  The revenue of the $\ratio$-markup mechanisms $\mech_{\ratio}$ on
 the triangle distribution $\tri_{\qo}$, for $\ratio \in (1,\infty)$ and
  $\qo \in [0,1)$, is
$$
\mecha_{\ratio}(\tri_{\qo}) 
= 
\frac{2\ratio}{(1-\qo)(\ratio-1)} 
\left(
\frac{1-\qo}{1-\qo+\qo\ratio}
+
\frac{\ln \left(
\frac{\ratio}{1-\qo+\qo\ratio}
\right)}{1-\ratio}
\right). 
$$
\end{lemma}

The proof of \Cref{clm:rev of r} is straightforward and is given in
\Cref{app:prior-independent}.
These lemmas allow us to numerically compute the expected revenues and approximation ratios of stochastic markup mechanisms given triangular distributions, 
which are illustrated in \cref{f:rev-approx}. 

\paragraph{Optimal Stochastic Markup Mechanism}
The following theorem characterizes the optimal
stochastic markup mechanism that is robust to scale against triangle distributions. 
The parameters of this optimal mechanism are the solution to \ a transcendental
expression (cf.\ \Cref{clm:rev of r}) that we are unable to solve
analytically.  Our proof will instead combine numeric calculations of
select points in parameter space with theoretical analysis to rule out
most of the parameter space.  For the remaining parameter space, we
can show that the expression is well-behaved and, thus, numeric
calculation can identify near optimal parameters.  Discussion of this
hybrid numerical and theoretical analysis can be found in
\Cref{app:prior-independent}.

\begin{theorem}\label{thm:triangle}

In the auxiliary markup--triangle game for a two-agent, single-item
environment, the designer's optimal strategy is

$\mechaARO$, which randomizes over the second-price auction
$\mecha_1$ with probability $\optweight$ and $\optratio$-markup
mechanism $\mecha_{\optratio}$ with probability $1-\optweight$,
where $\optweight \approx \asimp$ and $\optratio\approx
\rsimp$. The worst-case distribution for this mechanism is the
triangle distribution $\tri_{\optqo}$ with $\optqo \approx \qsimp$, and its approximation ratio is $\beta \approx \apxsimp$.
At this triangle, $\mecha_r(\tri_{\optqo})\leq1$ for every
markup $r\geq1$.
\end{theorem}
\begin{figure}[t]
  \begin{flushleft}
    \begin{minipage}[t]{0.48\textwidth}
      \centering
      \newcommand{\APXSCALE}{0.7}

\begin{tikzpicture}[scale = \APXSCALE]

\draw (-0.2,0) -- (9.5, 0);
\draw (0, -0.2) -- (0, 4.5);

\draw [dotted] (0, 0.9) -- (9, 0.9);
\draw [dotted] (9, 0) -- (9, 0.9);

\draw [dotted] (0.838, 0) -- (0.838, 1.72);

\begin{scope}[thick]
\draw (0, 1.8) -- (9, 0.9);

\draw plot [smooth, tension=0.8] coordinates {
(0, 0.9)
(0.18, 1.3136282419623115)
(0.36, 1.4640131820450633)
(0.54, 1.5748078501498688)
(0.72, 1.664835789425897)
(0.9, 1.7413094575111476)
(1.08, 1.8079002132369764)
(1.2600000000000002, 1.8667921201394844)
(1.44, 1.9194246152090024)
(1.6199999999999999, 1.9667793471684476)
(1.8, 2.0095760109323155)
(1.98, 2.04837544425269)
(2.16, 2.083615029815598)
(2.34, 2.116216216216216)
(2.5200000000000005, 2.15)
(2.6999999999999997, 2.185714285714286)
(2.88, 2.223529411764706)
(3.06, 2.2636363636363637)
(3.2399999999999998, 2.30625)
(3.42, 2.351612903225807)
(3.6, 2.4000000000000004)
(3.78, 2.4517241379310346)
(3.96, 2.507142857142857)
(4.140000000000001, 2.5666666666666664)
(4.32, 2.6307692307692307)
(4.5, 2.7)
(4.68, 2.7750000000000004)
(4.86, 2.856521739130435)
(5.040000000000001, 2.9454545454545458)
(5.22, 3.0428571428571427)
(5.3999999999999995, 3.1499999999999995)
(5.58, 3.2684210526315787)
(5.76, 3.4)
(5.94, 3.5470588235294116)
(6.12, 3.7125)
(6.3, 3.9)
(6.4799999999999995, 4.114285714285714)
(6.66, 4.361538461538461)
(6.84, 4.65)
(7.0200000000000005, 4.990909090909091)
};

\end{scope}

\draw (0, -0.5) node {$0$};
\draw (9, -0.5) node {$1$};
\draw (1, -0.5) node {$0.093$};
\draw (-0.4, 0) node {$0$};
\draw (-0.4, 0.9) node {$1$};
\draw (-0.4, 1.8) node {$2$};

\end{tikzpicture}
    \end{minipage}    
    \begin{minipage}[t]{0.48\textwidth}
      \centering
      \newcommand{\MKUPSCALE}{0.7}

\begin{tikzpicture}[scale = \MKUPSCALE]

\draw (-0.2,0) -- (9.5, 0);
\draw (0, -0.2) -- (0, 4.5);

\draw [dotted] (9, 0) -- (9, 1.8318570061324486);
\draw [dotted] (0, 4) -- (2.52, 4);
\draw [dotted] (2.52, 0) -- (2.52, 4);

\draw[thick] (0,2.1378860000000017) circle (0.1cm);
\fill[black] (0,4) circle (0.1cm);

\begin{scope}[thick]

\draw plot [smooth, tension=0.8] coordinates {
(0, 2.1378860000000017)
(0.18000000000000016, 2.4858546236856434)
(0.35999999999999993, 2.7780923762522782)
(0.5400000000000001, 3.0236057552038424)
(0.7199999999999999, 3.2295397242542947)
(0.9, 3.401652011704517)
(1.0800000000000003, 3.54464523357451)
(1.26, 3.662405252282408)
(1.4400000000000002, 3.758175921484643)
(1.6199999999999999, 3.8346895991501135)
(1.8, 3.894266236323606)
(1.9800000000000002, 3.9388897117643253)
(2.1600000000000006, 3.9702674089368877)
(2.34, 3.9898772626667665)
(2.52, 3.9990053071619727)
(2.7, 3.9987759336185498)
(2.8800000000000003, 3.9901764886222963)
(3.0600000000000005, 3.974077433896767)
(3.2399999999999998, 3.951248991475193)
(3.42, 3.922374981523067)
(3.6, 3.8880643995024613)
(3.7800000000000002, 3.8488611591964172)
(3.9600000000000004, 3.8052523372212086)
(4.14, 3.7576751852487114)
(4.32, 3.706523122688221)
(4.5, 3.6521508810329983)
(4.680000000000001, 3.594878938548881)
(4.86, 3.534997358325196)
(5.04, 3.472769122332881)
(5.22, 3.408433037845761)
(5.4, 3.3422062794818572)
(5.579999999999999, 3.2742866195242932)
(5.760000000000001, 3.204854390562332)
(5.9399999999999995, 3.1340742174459386)
(6.120000000000001, 3.0620965497569976)
(6.3, 2.989059021220733)
(6.4799999999999995, 2.915087658517148)
(6.66, 2.8402979586524637)
(6.84, 2.76479585129049)
(7.0200000000000005, 2.688678560128462)
(7.2, 2.6120353754500805)
(7.38, 2.534948348339359)
(7.5600000000000005, 2.4574929156391434)
(7.74, 2.3797384635475645)
(7.920000000000001, 2.3017488367286383)
(8.1, 2.223582798942793)
(8.28, 2.1452944504553884)
(8.46, 2.0669336068375266)
(8.64, 1.9885461432174623)
(8.82, 1.9101743075598314)
(9.0, 1.8318570061324486)
};

\end{scope}

\draw (0, -0.5) node {$1$};
\draw (9, -0.5) node {$5$};
\draw (2.52, -0.5) node {$2.45$};
\draw (-0.7, 0) node {$0.8$};
\draw (-0.7, 2.137) node {$0.91$};
\draw (-0.4, 4) node {$1$};

\end{tikzpicture}
    \end{minipage}
    \vspace{-18pt}
  \end{flushleft}
  \caption{\label{f:rev-approx} The figure on the left plots, as a function of $\qo$, the
    approximation ratio $\APXSPA(\qo)$ of the second-price auction
    $\mecha_1$ against triangle distribution $\tri_{\qo}$ (straight line), and the approximation ratio $\APXR(\qo)$ of the optimal non-trivial
    markup mechanism against triangle distribution $\tri_{\qo}$ (curved line).  These functions cross at $\optqo = \qval$.  The figure on the
    right plots the revenue of the $\ratio$ markup mechanism
    $\mech_{\ratio}$ on triangle distribution $\tri_{\optqo}$ as a
    function of markup $\ratio$, i.e.,
    $\mech_{\ratio}(\tri_{\optqo})$.  Notice that, by choice of $\optqo$, the optimal
    non-trivial markup mechanism has the same revenue as the
    second-price auction.}
\end{figure}
Intuitively, the auxiliary markup--triangle problem can be viewed as a zero-sum game between the designer and an adversary, 
where the designer chooses a stochastic markup mechanism, 
the adversary chooses a triangle distribution (and its induced revenue curve), 
and the payoff of the designer is the approximation ratio $\OPT_{\dist}(\dist)/\mecha(\dist)$ (see \Cref{def:pimd}).
The solution to this game is characterized by a Nash equilibrium between the designer and the adversary.

The high level approach of this proof is to identify the triangle
$\tri_{\optqo}$ for which the designer is indifferent between the
second price auction $\mech_1$ and the optimal (non-trivial) markup
mechanism, denoted $\mech_{\optratio}$.  For such a distribution
$\tri_{\optqo}$, the designer is also indifferent (in minimizing the
approximation ratio) between any mixture over $\mech_1$ (with
probability~$\secprob$) and $\mech_{\optratio}$ (with probability
$1-\secprob$), and all other $\ratio$-markup mechanisms for $\ratio
\not \in \{1,\optratio\}$ are inferior (see \Cref{f:rev-approx}).  We
then identify the $\optweight$ for which the adversary's best
response (in maximizing the approximation ratio) to $\mechaARO$ is
the distribution $\tri_{\optqo}$. 
Thus, $\mechaARO$ and $\tri_{\optqo}$ form a Nash equilibrium of the
auxiliary markup--triangle game.
The parameters can be numerically identified as
$\optweight \approx \aval$, $\optratio\approx\rval$, $\optqo \approx
\qval$, and the approximation ratio is $\piratio \approx \apxval$.

\subsection{Markup Reduction and Worst-Case Triangle Distributions}
  \label{sec:best response}

We first reduce the robust lower-bound problem for highest-value-only mechanisms
to stochastic markups on atomless distributions. We then show that
nature has a best response among truncated distributions to each
relevant stochastic markup mechanism and that triangle distributions
are best among the truncated distributions.

Triangle distributions are known to be worst case for other questions
of interest in mechanism design, e.g., approximation by anonymous
reserves and anonymous pricings \citep{AHNPY-18}.  The proof that
triangle distributions are worst-case for two-agent revenue maximization under the scale-robust analysis framework
is significantly more involved than these previous results.

\subsubsection{Atomless Reduction to Stochastic Markups}

\begin{lemma}[Atomless Markup Reduction]
\label{lem:perturbed-reduction}
In the i.i.d., two-agent, single-item environment, every feasible,
DSIC, ex-post IR, scale-invariant, highest-value-only mechanism has worst-case
approximation ratio over $\regular$ at least $\beta$, the value in
\Cref{thm:triangle}. More precisely, after payment normalization and
anonymization, the mechanism is weakly dominated on every atomless
i.i.d.\ distribution by a stochastic markup mechanism. Atomless
regular perturbations of $\tri_{\optqo}$ then witness the lower bound
$\beta$.
\end{lemma}

\begin{proof}
Let $M$ be such a mechanism, and let $\bar M=(x,p)$ be obtained by
normalizing payments so that $p_i(0,\valsmi)=0$ and then anonymizing
by randomly permuting the bidders, as in the proof of
\Cref{thm:scale_invariant_is_opt}. Normalization weakly raises revenue,
while anonymization preserves revenue on every i.i.d.\ distribution
as well as all the maintained mechanism properties. Thus
\[
 \sup_{\dist\in\regular}\frac{\OPT_{\dist}(\dist)}{M(\dist)}
 \geq
 \sup_{\dist\in\regular}\frac{\OPT_{\dist}(\dist)}{\bar M(\dist)}.
\]
It is therefore enough to prove the lower bound for $\bar M$, which
we relabel $M$.

For $r>1$, define the allocation to the strictly higher bidder by
\[
 g(r):=x_1(r,1).
\]
Scale invariance implies that the high bidder's allocation at every
strict profile $(h,\ell)$ with $h>\ell>0$ is $g(h/\ell)$.
Under the highest-value-only restriction, the lower bidder's allocation is zero, and DSIC makes
$g$ nondecreasing. The normalized payment identity gives, with
$r=h/\ell$,
\begin{equation}\label{eq:off-tie-payment}
 p_1(h,\ell)=\ell\left(rg(r)-\int_1^r g(s)\,ds\right),
 \qquad p_2(h,\ell)=0.
\end{equation}
By anonymity, the same formula holds after interchanging the bidders.

Let $g^+(r)=\lim_{s\downarrow r,\,s>r}g(s)$, including
$g^+(1)=\lim_{s\downarrow1}g(s)$, and put
$\gamma=\lim_{r\to\infty}g(r)\leq1$. The right-continuous
nondecreasing function $g^+$ is the cumulative distribution function
of a finite Stieltjes measure $\mu$ on $[1,\infty)$, defined by
$\mu([1,r])=g^+(r)$ and having total mass $\gamma$. The functions $g$
and $g^+$ differ only at the countable
set of discontinuities of $g$, so their Lebesgue integrals agree. At
every continuity point of $g$, integration by parts therefore gives
\[
 \int_{[1,r]}s\,d\mu(s)
 =rg(r)-\int_1^r g(s)\,ds.
\]
Complete $\mu$ to a probability measure by putting its missing mass at
markup one:
\[
 \widehat\mu:=\mu+(1-\gamma)\delta_1,
\]
and let $\widehat M$ be the corresponding stochastic markup
mechanism. Equation~\eqref{eq:off-tie-payment} shows that
$\widehat M$ has weakly higher revenue than $M$ at every strict
profile whose bid ratio is a continuity point of $g$.

A monotone function has only countably many discontinuities. If
$\dist$ is atomless, i.i.d.\ draws tie with probability zero and their
ratio lies in that countable exceptional set with probability zero.
Indeed, for every fixed $c>0$,
$\Pr[V_1=cV_2]=\int\Pr[V_1=cv]\,d\dist(v)=0$; also
$\Pr[V_i=0]=0$, so the ratio is defined almost surely.
Consequently,
\begin{equation}\label{eq:atomless-domination}
 M(\dist)\leq \widehat M(\dist)
 \qquad\text{for every atomless i.i.d.\ distribution }\dist.
\end{equation}

We next approximate the adversarial triangle by atomless distributions
in $\regular$. Put $q=\optqo\in(0,1)$. For
$0<\varepsilon<1/q^2$, define a distribution $\dist_{q,\varepsilon}$
by its price-posting revenue curve
\begin{equation}\label{eq:smoothed-triangle}
 R_{q,\varepsilon}(u)=
 \begin{cases}
 \dfrac{u}{q}+\varepsilon u(q-u),&0\leq u\leq q,\\[4pt]
 \dfrac{1-u}{1-q},&q\leq u\leq1.
 \end{cases}
\end{equation}
The first piece has second derivative $-2\varepsilon$ and left
derivative $1/q-\varepsilon q>0$ at $q$, whereas the second piece has
derivative $-1/(1-q)$. Hence the curve is concave, is maximized at
$q$, and has normalized monopoly revenue one. For $0<u\leq q$, its
quantile-value function is
\[
 v_{q,\varepsilon}(u)=\frac{R_{q,\varepsilon}(u)}{u}
 =\frac1q+\varepsilon(q-u),
 \qquad 0<u\leq q,
\]
and it agrees with $\tri_q$ for $u\geq q$. Thus the quantile-value
function, whose second piece is
$v_{q,\varepsilon}(u)=(1-u)/(u(1-q))$, is continuous and strictly
decreasing. Hence $\dist_{q,\varepsilon}$
is atomless and regular. It also has bounded support, and therefore a
finite first moment. Notice that it is not itself truncated; it is
nevertheless an admissible member of $\regular$.

By the revenue-curve formula in \Cref{l:DRY-15}, the smallest monotone
concave upper bound equals $R_{q,\varepsilon}$ on $[0,q]$ and the
constant $1$ on
$[q,1]$. Therefore
\begin{equation}\label{eq:smoothed-opt}
 \OPT_{\dist_{q,\varepsilon}}(\dist_{q,\varepsilon})
 =2\left(\int_0^qR_{q,\varepsilon}(u)\,du+1-q\right)
 =2-q+\frac{\varepsilon q^3}{3}
 \longrightarrow 2-q.
\end{equation}

For completeness, couple $\dist_{q,\varepsilon}$ and $\tri_q$ by
common quantiles. Their values converge uniformly as
$\varepsilon\downarrow0$. For a fixed markup $r\geq1$, its realized
payment is
\[
 r\min\{v_1,v_2\}
 \boldsymbol 1\!\left\{\max\{v_1,v_2\}
       \geq r\min\{v_1,v_2\}\right\}.
\]
For $r=1$, the payment is $\min\{v_1,v_2\}$ and is continuous. For
every fixed $r>1$, the event
$\max\{V_1,V_2\}=r\min\{V_1,V_2\}$ has probability zero under
$\tri_q^{\otimes2}$: the sole atom of the bid ratio is at one. Thus
the payments converge almost surely. For all sufficiently small
$\varepsilon$, they are bounded by a common constant
$1/q+\varepsilon_0q$. Joint dominated convergence over values and
$r\sim\widehat\mu$ yields
\begin{equation}\label{eq:markup-smoothing-limit}
 \widehat M(\dist_{q,\varepsilon})
 \longrightarrow \widehat M(\tri_q).
\end{equation}

In the proof of \Cref{thm:triangle}, the defining equality
$\APXR(q)=\APXSPA(q)$ and the definition
$R(q)=\sup_{r>1}\mecha_r(\tri_q)$ imply $R(q)=1$.
Together with \Cref{lem:spa rev}, this gives
$\mecha_r(\tri_q)\leq1$ for every $r\geq1$, as recorded in the
theorem statement. Hence $\widehat M(\tri_q)\leq1$. If this last
revenue is zero, the desired approximation ratios diverge; otherwise,
combining
\eqref{eq:atomless-domination}--\eqref{eq:markup-smoothing-limit},
\[
 \sup_{\dist\in\regular}
 \frac{\OPT_{\dist}(\dist)}{M(\dist)}
 \geq
 \liminf_{\varepsilon\downarrow0}
 \frac{\OPT_{\dist_{q,\varepsilon}}(\dist_{q,\varepsilon})}
      {M(\dist_{q,\varepsilon})}
 \geq 2-q=\beta.
\]
This proves the claimed robust lower bound.
\end{proof}

The domination in \eqref{eq:atomless-domination} need not hold on the
atomic triangle itself, because tie behavior may give
$M(\tri_q)>\widehat M(\tri_q)$. The lower bound uses the admissible
atomless distributions $\dist_{q,\varepsilon}$, on which exact ties
have probability zero. Thus atomless approximation, rather than
pointwise domination on every truncated distribution, is the property
required for the robust minimax argument.

\subsubsection{Best Response of Triangle Distributions}

Next we will give a sequence of results that culminate in the
observation that for any regular distribution and any stochastic
markup mechanism with probability $\secprob$ at least $\sfrac 2 3$ on
the second-price auction (as does $\mechaARO$, characterized in
\Cref{thm:triangle}), either the triangulation of the distribution or
the point mass $\tri_{1}$ has (weakly) higher approximation ratio.  As
the notation indicates, the point mass distribution $\tri_{1}$ is a
triangle distribution.

\begin{lemma}\label{thm:tri}
  For i.i.d., two-agent, single-item environments and any regular
  distribution $\dist$ and any stochastic markup mechanism $\mech$
  that places probability $\alpha \in [\sfrac 2 3,1]$ on the
  second-price auction, either the triangulation of the distribution
  $\distTri$ or the point mass $\tri_{1}$ has (weakly) higher
  approximation ratio.  I.e.,
  $\max\left\{\frac{\optf{\distTri}}{\mech(\distTri)},\frac{\optf{\tri_1}}{\mech(\tri_1)}\right\}
  \geq \frac{\optf{\dist}}{\mech(\dist)}$.
  \end{lemma}

To prove this lemma we give a sequence of results showing that for
any regular distribution, a corresponding truncated distribution is only
worse; for any truncated distribution and a fixed stochastic markup
mechanism (that mixes over $\mech_1$ and some $\mech_{\ratio}$), a
corresponding quadrilateral distribution (based on $\ratio$) is only
worse; and for any quadrilateral distribution, a
corresponding triangle distribution (independent of $\ratio$) is only
worse.  The theorem follows from combining these results.  The first
step assumes that the probability that the stochastic markup mechanism
places on the second price auction is $\secprob \in [\sfrac 1 2,1]$;
the last step further assumes that $\secprob \in [\sfrac 2 3, 1]$.

\paragraph{Best response of truncated distributions}
To begin, the following lemma shows that the best response of the
adversary to a relevant stochastic markup mechanism is a truncated
distribution.  Recall that by \citet{FILS-15} the optimal scale-robust mechanism is strictly better than a 2-approximation.  On the
other hand, any stochastic markup mechanism that places probability
$\secprob$ on the second-price auction $\mech_1$ has approximation ratio at least $\sfrac 1 {\secprob}$. Specifically, on the
(degenerate) distribution that places all probability mass on 1,
a.k.a.\ $\tri_{1}$, the approximation factor of such a stochastic
markup mechanism is exactly $\sfrac 1 {\secprob}$.  We conclude that
all relevant stochastic markup mechanisms place probability $\secprob
> \sfrac 1 2$ on the second-price auction.  Thus, this lemma applies
to all relevant mechanisms.

\begin{lemma}\label{lem:truncate}
  For i.i.d., two-agent, single-item environments, any regular
  distribution $\dist$, and any stochastic markup mechanism $\mech$
  that places probability $\alpha \in [\sfrac 1 2,1]$ on the
  second-price auction; either the truncation of the distribution
  $\dist'$ or the point mass distribution $\tri_1$ has (weakly) higher
  approximation ratio.  I.e.,
  $\max\left\{\frac{\optf{\dist'}}{\mech(\dist')},\frac{\optf{\tri_1}}{\mech(\tri_1)}\right\}
  \geq \frac{\optf{\dist}}{\mech(\dist)}$.
\end{lemma}

\begin{figure}[t]
\begin{flushleft}
\begin{minipage}[t]{0.48\textwidth}
\centering
\newcommand{\REVONESCALE}{0.7}

\begin{tikzpicture}[scale = \REVONESCALE]

\draw [name path global = A] plot [smooth, tension=0.45] coordinates {(0, 0) (0.5, 2.5) (1,3.5) (1.9, 4.4) (2.5, 4.8) (3, 5)};

\draw [dotted] (0, 5) -- (3, 5);
\draw [name path global = B, dotted] (3, 0) -- (3, 5);

\tikzfillbetween[of=A and B]{pattern=north east lines, pattern color=gray};

\fill[color=gray!40!white] (3, 0.01) -- (3, 5) -- (9, 5) -- (9, 0.01);

\draw plot [smooth, tension=0.5] coordinates {(3, 5) (4.5, 4.76) (6, 4) (8, 2) (9, 0)};

\draw (-0.5, 5) node {$1$};

\draw (0, -0.5) node {$0$};
\draw (9, -0.5) node {$1$};
\draw (3, -0.5) node {$\qo$};

\draw (-0.2,0) -- (10, 0);
\draw (0, -0.2) -- (0, 5.5);

\end{tikzpicture}
\end{minipage}
\begin{minipage}[t]{0.48\textwidth}
\centering
\newcommand{\REVTWOSCALE}{0.7}

\begin{tikzpicture}[scale = \REVTWOSCALE]

\draw (-0.2,0) -- (10, 0);
\draw (0, -0.2) -- (0, 5.5);

\draw (0, 0) -- (3, 5);

\draw [dotted] (0, 5) -- (3, 5);
\draw [dotted] (3, 0) -- (3, 5);

\draw [name path global = B] (3, 0.01) -- (9, 0.01);

\draw [name path global = A] plot [smooth, tension=0.5] coordinates {(3, 5) (4.5, 4.76) (6, 4) (8, 2) (9, 0)};

\fill[pattern=north east lines, pattern color=gray] (0, 0) -- (3, 5) -- (3, 0);

\tikzfillbetween[of=A and B]{gray!40!white};

\draw plot [smooth, tension=0.5] coordinates {(3, 5) (4.5, 4.76) (6, 4) (8, 2) (9, 0)};

\draw (-0.5, 5) node {$1$};

\draw (0, -0.5) node {$0$};
\draw (9, -0.5) node {$1$};
\draw (3, -0.5) node {$\qo$};

\end{tikzpicture}
\end{minipage}
\vspace{-18pt}
\end{flushleft}
\caption{\label{f:rev-decompose} The illustration of the revenue
  decomposition of \Cref{lem:truncate} for $\mecha$ on distribution
  $\dist$ and truncation $\dist'$ for the optimal mechanism and
  second-price auction.  The thin black line on the left and right
  figures are the revenue curves corresponding to $\dist$ and
  $\dist'$, respectively.  The dashed area on the left represents
  $\OPT_+ = \SPA_+$ and the gray area on the left represents $\OPT_- =
  \OPT'_-$.  The dashed area on the right represents $\OPT'_+ = \SPA'_+$
  and the
  gray area on the right represents $\SPA'_- = \SPA_-$.}
\end{figure}

\begin{proof}
  It can be assumed that the approximation of stochastic markup
  mechanism $\mecha$ on distribution $\dist$ is at least $\sfrac 1
  {\secprob}$ (where $\secprob$ denotes the probability that $\mecha$
  places on the second-price auction).  Notice that the revenue
  $\mecha$ on the point mass on 1 (a truncated distribution) is
  $\secprob$ and the optimal revenue on this distribution is 1.  If
  the approximation factor $\sfrac{\optf{\dist}}{\mecha(\dist)}$ is
  less than $\sfrac{1}{\secprob}$ then the point mass on 1 (a truncated
  distribution) achieves a higher approximation than $\dist$ and the
  lemma follows.  For the remainder of the proof, assume that the
  approximation factor of mechanism $\mecha$ on distribution $\dist$
  is more than $\sfrac{1}{\secprob}$.

  View the stochastic markup mechanism $\mecha$ as a distribution over two
  mechanisms: the second-price auction $\mech_1$ with probability
  $\alpha$, and $\mech_{*}$, a distribution over non-trivial markup
  mechanisms $\mech_{\ratio}$ with $\ratio > 1$, with probability
  $1-\alpha$.  The optimal mechanism is $\OPT_{\dist}$.  Decompose the
  revenue from distribution $\dist$ across these three mechanisms as
  follows.  Denote the monopoly quantile of $\dist$ by $\qo$.
  See \Cref{f:rev-decompose}.

  \begin{itemize}
  \item $\OPT_+$ and $\OPT_-$ give the expected revenue of the
    optimal mechanism from each agent when their opponent has value above and below the
    monopoly price $\valf_{\dist}(\qo)$.
    
  \item $\SPA_+ = \OPT_+$ and $\SPA_-$ give the expected revenue of
    the second-price auction $\mecha_1$ from each agent when their opponent has value above
    and below the monopoly price.
    
  \item $\MKUP_+$ and $\MKUP_-$ give the expected revenue of the
    stochastic markup mechanism~$\mecha_*$ when the realized prices are (strictly)
    above and (weakly) below the monopoly price.
  \end{itemize}
Consider truncating the distribution $\dist$ at the monopoly quantile
$\qo$ to obtain $\dist' \in \TRUNCF$.  Define analogous quantities (with identities):
\begin{itemize}
\item $\OPT'_+ \leq \OPT_+$ and $\OPT'_- = \OPT_-$.

  Identities follow from the geometric analysis of \Cref{l:DRY-15}.
  
\item $\SPA'_+ = \OPT'_+$ and $\SPA'_- = \SPA_-$.

  Identities follow from the geometric analysis of \Cref{l:DRY-15}.
  
\item $\MKUP'_+ = 0$ and $\MKUP'_- = \MKUP_-$.

  Values above the monopoly price are not supported by the truncated
  distribution, so the revenue from those prices is zero.  On the
  other hand, prices (weakly) below the monopoly price are bought with
  the exact same probability as the cumulative distribution function
  $\dist'$ and $\dist$ are the same for these prices.
 
\end{itemize}
The remainder of the proof follows a straightforward calculation.
Write the approximation ratio of $\mecha$ on distribution $\dist$
(using the given identities) and rearrange:
\begin{align*}
\frac{\optf{\dist}}{\mecha(\dist)}
&= \frac{\OPT_+ + \OPT_-}{\secprob\, (\OPT_+ + \SPA_-) + (1-\secprob)\,(\MKUP_+ + \MKUP_-)}\\
&= \frac{\OPT_+ + \left[\OPT_-\right]}{\secprob\, \OPT_+  + \left[\secprob \SPA_- + (1-\secprob)\,(\MKUP_+ + \MKUP_-)\right]}\\
\intertext{Since the approximation ratio on $\dist$ is at least $\sfrac 1 \secprob$, the ratio of the first term in the numerator and denominator is at most the ratio of the remaining terms [in brackets]:}
\frac{1}{\secprob} &= \frac{\OPT_+}{\secprob\,\OPT_+} \leq \frac{\left[\OPT_-\right]}{\left[\secprob \SPA_- + (1-\secprob)\,(\MKUP_+ + \MKUP_-)\right]}\\
\intertext{Now write the approximation ratio of $\mecha$ on truncation $\dist'$ (using the given identities) and bound:}
\frac{\optf{\dist'}}{\mecha(\dist')}
&= \frac{\OPT'_+ + \left[\OPT_-\right]}{\secprob\, \OPT'_+ + \left[\secprob\, \SPA_- + (1-\secprob)\,\MKUP_-\right]}\\
&\geq \frac{\OPT'_+ + \left[\OPT_-\right]}{\secprob\, \OPT'_+ + \left[\secprob\, \SPA_- + (1-\secprob)\,(\MKUP_+ + \MKUP_-)\right]}\\
&\geq \frac{\OPT_+ + \left[\OPT_-\right]}{\secprob\, \OPT_+ + \left[\secprob\, \SPA_- + (1-\secprob)\,(\MKUP_+ + \MKUP_-)\right]}\\
&= \frac{\optf{\dist}}{\mecha(\dist)}.
\end{align*}
The calculation shows that, for any distribution $\dist$, the
truncated distribution $\dist'$ increases the approximation factor of
the stochastic markup mechanism.  Thus, the worst-case distribution is
truncated.
\end{proof}        

\begin{figure}[t]
\begin{flushleft}
\begin{minipage}[t]{0.48\textwidth}
\centering
\newcommand{\QONESCALE}{0.7}

\begin{tikzpicture}[scale = \QONESCALE]

\draw (-0.2,0) -- (10, 0);
\draw (0, -0.2) -- (0, 5.5);

\draw (0, 0) -- (3, 5);
\draw plot [smooth, tension=0.6] coordinates {(6, 4) (7, 3) (9, 0)};

\draw plot [smooth, tension=0.6] coordinates {(3, 5) (4.5, 4.7) (6, 4)};


\draw [dotted] (0, 5) -- (3, 5);
\draw [dotted] (0, 0) -- (7.2, 4.8);
\draw [dotted] (3, 0) -- (3, 5);

\draw [dotted] (6, 0) -- (6, 4);

\begin{scope}[very thick]
\draw [dashed, gray] (0, 0) -- (3, 5);
\draw [dashed, gray] (3, 5) -- (6, 4);
\draw [dashed, gray] plot [smooth, tension=0.6] coordinates {(6, 4) (7, 3) (9, 0)};
\end{scope}

\draw (-0.5, 5) node {$1$};

\draw (0, -0.5) node {$0$};
\draw (9, -0.5) node {$1$};
\draw (3, -0.5) node {$\qo$};
\draw (6, -0.5) node {$\qt$};
\draw (6.5, 5) node {$\sfrac{1}{r\qo}$};

\end{tikzpicture}
\end{minipage}
\begin{minipage}[t]{0.48\textwidth}
\centering
\newcommand{\QTWOSCALE}{0.7}

\begin{tikzpicture}[scale = \QTWOSCALE]

\draw (-0.2,0) -- (10, 0);
\draw (0, -0.2) -- (0, 5.5);

\draw (0, 0) -- (3, 5);
\draw plot [smooth, tension=0.6] coordinates {(6, 4) (7, 3) (9, 0)};

\draw (3, 5) -- (6, 4);

\draw [dotted] (0, 5) -- (3, 5);
\draw [dotted] (0, 0) -- (7.2, 4.8);
\draw [dotted] (3, 0) -- (3, 5);

\draw [dotted] (6, 0) -- (6, 4);

\begin{scope}[very thick]
\draw [dashed, gray] (0, 0) -- (3, 5);
\draw [dashed, gray] (3, 5) -- (6, 4);
\draw [dashed, gray] (6, 4) -- (9, 0);
\end{scope}

\draw (-0.5, 5) node {$1$};

\draw (0, -0.5) node {$0$};
\draw (9, -0.5) node {$1$};
\draw (3, -0.5) node {$\qo$};
\draw (6, -0.5) node {$\qt$};
\draw (6.5, 5) node {$\sfrac{1}{r\qo}$};

\end{tikzpicture}
\end{minipage}
\vspace{-18pt}
\end{flushleft}
\caption{\label{f:proof-quad} The main two steps of \Cref{lem:quad}
  are illustrated.  In the first step (right-hand side), the revenue
  curves of distributions $\distTrunc$ (thin, solid, black) and
  $\dist\primed$ (thick, dashed, gray) are depicted.  In the second
  step, the revenue curves of the distributions $\dist\primed$ (thin,
  solid, black) and $\distQr$ (thick, dashed, gray) are depicted.  In
  both cases the revenue of the $\ratio$-markup mechanism is higher
  on the thin, solid, black curve than the thick, dashed, gray curve.}
\end{figure}

\paragraph{Best response of quadrilateral distributions}
The next step is to show that, among truncated distributions, the
worst-case distributions for stochastic markup mechanisms are those
with quadrilateral-shaped revenue curves, i.e., ones that are
piecewise linear with three pieces (see \Cref{f:quad}).  Recall that
for a truncated distribution at the monopoly quantile $\qo$, the upper
bound of the support is a point mass at $\sfrac 1 \qo$.

\begin{definition}[Quadrilateral Distribution]
\label{def:quadrilateral}

A \emph{normalized quadrilateral distribution} with parameters
$\qo\in(0,1]$, $\qt$, and $\ratio>1$, where
\[
 \frac{\qo\ratio}{\qo\ratio+1-\qo}
 \leq\qt\leq\min\{\ratio\qo,1\},
\]
denoted by $\Qr_{\qo,\qt,\ratio}$, is defined by
\[
\quantf_{\Qr_{\qo,\qt,\ratio}}(\val)=
\begin{cases}
\dfrac{\qt}{\qt+\val\ratio\qo(1-\qt)},
 & \val<\dfrac1{\ratio\qo},\\[5pt]
\dfrac{\qt\qo(\ratio-1)}
{\val\ratio\qo(\qt-\qo)+(\ratio\qo-\qt)},
 & \dfrac1{\ratio\qo}\leq\val\leq\dfrac1\qo,\\[5pt]
0, & \val>\dfrac1\qo.
\end{cases}
\]

\end{definition}

\noindent The following lemma summarizes an analysis from \citet{AB-18} and is
useful in bounding the revenue from markup mechanisms.

\begin{lemma}[\citealp{AB-18}]
  \label{l:AB-18}
Consider the $\ratio$-markup mechanism, two i.i.d.\ regular agents
with value distribution $\dist$, quantile $\qt$ corresponding to the
monopoly price divided by~$\ratio$, and the distribution
$\Tilde{\dist}$ that corresponds to $\dist$ ironed on $[\qt,1]$: the
virtual surplus from quantiles $[\qt,1]$ is higher for $\dist$ than
for $\Tilde{\dist}$.
\end{lemma}

\begin{proof}
  The proof of this lemma is technical and non-trivial.  It is given in 
  the proof of Proposition~4 of \citet{AB-18}.
\end{proof}

The next lemma reduces the worst case distribution from the family of truncated distributions to the family of quadrilateral distributions.   The reduction is illustrated in \Cref{f:proof-quad}, by showing that ironing the revenue curves sequentially within $[\qo,\qt]$ and $[\qt, 1]$ decreases the revenue of the stochastic markup mechanism.  The optimal revenue is not affected because it is obtained using a reserve price corresponding to the monopoly quantile $\qo$ and it is agnostic to the shape of the revenue curve for $\quant> \qo$.

\begin{lemma}\label{lem:quad}
    For i.i.d., two-agent, single-item environments, any truncated
    distribution $\distTrunc$, and any stochastic markup mechanism
    $\mechaAR$ with probability $\secprob$ on the second-price auction
    $\mech_1$ and probability $1-\secprob$ on non-trivial markup
    mechanism $\mech_r$; there is a quadrilateral distribution
    $\distQr$ with the same optimal revenue and (weakly) lower revenue
    in $\mechaAR$.  I.e., $\optf{\distQr} = \optf{\distTrunc}$ and
    $\mechaAR(\distQr) \leq \mechaAR(\distTrunc)$.
\end{lemma}

\begin{proof}
  On any normalized truncated distribution with monopoly quantile
  $\qo$, the optimal revenue is $2-\qo$ (\Cref{lem:opt rev}).  Thus,
  to prove the lemma it is sufficient to show that for any truncated
  distribution $\distTrunc \in \TRUNCF$ with monopoly quantile $\qo$
  there is a normalized quadrilateral distribution $\distQr \in \QRF
  \subset \TRUNCF$ with monopoly quantile $\qo$ and lower revenue in
  $\mechaAR$.

  The quadrilateral distribution $\distQr$ is obtained by ironing
  $\distTrunc$ on $[\qo,\qt]$ and $[\qt,1]$ where quantile $\qt$
  satisfies $\valf_{\distTrunc}(\qo) =
  \ratio\,\valf_{\distTrunc}(\qt)$.  We consider an intermediary
  distribution $\dist\primed$ that is $\distTrunc$ ironed only on
  $[\qo,\qt]$.  See \Cref{f:proof-quad}.  The proof approach is to
  show that $\mechaAR(\distTrunc) \geq \mechaAR(\dist\primed) \geq
  \mechaAR(\distQr)$.

  As $\mechaAR$ is a convex combination of the second-price auction
  $\mecha_1$ and the $\ratio$-markup mechanism $\mecha_{\ratio}$.  It
  suffices to show the inequalities above hold for both auctions.  In fact,
  the result holds for the second-price auction from the geometric
  analysis of revenue of \Cref{l:DRY-15}.  The revenue of the
  second-price auction for two i.i.d.\ agents is twice the area under
  the revenue curve.  As the revenue curve has weakly smaller area
  from $\distTrunc$ to $\dist\primed$ to $\distQr$, we have
  $\mecha_1(\distTrunc) \geq \mecha_1(\dist\primed) \geq \mecha_1(\distQr)$.
  Below, we analyze the $\ratio$-markup mechanism $\mecha_{\ratio}$.

  The following price-based analysis shows that
  $\mecha_{\ratio}(\distTrunc) \geq \mech_{\ratio}(\dist\primed)$:
  \begin{itemize}
  \item The revenue from quantiles in $[0,\qo]$ is unchanged.
    
    These quantiles are offered prices from quantiles in $[\qt,1]$.
    The values of quantiles $[0,\qo]$ and $[\qt,1]$ are the same for
    both distributions; thus, the revenue is unchanged.
    
  \item The revenue from quantiles in $[\qo,\qt]$ decreases.

    These quantiles are offered prices from quantiles in $[\qt,1]$.
    For the distribution $\dist\primed$ relative to $\distTrunc$:
    Values are lower at any quantile $\quant \in [\qo,\qt]$; the
    distribution of prices (from quantiles in $[\qt,1]$) is the same.
    Thus, revenue is lower.

  \item The revenue from quantiles in $[\qt, 1]$ is unchanged.

    These quantiles are in $[\qt,1]$ and are offered prices from
    quantiles in $[\qt,1]$.  The distributions are the same for these
    quantiles; thus, the revenue is unchanged.
  \end{itemize}

  The following virtual-surplus-based analysis shows that
  $\mecha_{\ratio}(\dist\primed) \geq \mech_{\ratio}(\distQr)$:
  \begin{itemize}
  \item The virtual surplus of quantiles in $[0,\qo]$ is unchanged.
    
    These quantiles have the same virtual values under the two
    distributions and the same probability of winning, i.e., $1-\qt$
    (when the other agent's quantile is in $[\qt,1]$).
    
  \item The virtual surplus of quantiles in $[\qo,\qt]$ is decreased.

    Their prices come from quantiles in $[\qt,1]$ which are decreased;
    thus, their probabilities of winning are increased.  Their virtual
    values are negative, so these increased probabilities of winning
    result in decreased virtual surplus.

  \item The virtual surplus of quantiles in $[\qt,1]$ is decreased.

    This result is given by \Cref{l:AB-18}.  \qedhere
  \end{itemize}
\end{proof}

\begin{figure}[t]
\centering
\begin{tikzpicture}[scale = 0.7]

\draw (-0.2,0) -- (10, 0);
\draw (0, -0.2) -- (0, 5.5);

\draw (0, 0) -- (3, 5);
\draw (3, 5) -- (6, 4);
\draw (6, 4) -- (9, 0);

\draw [dotted] (0, 5) -- (3, 5);
\draw [dotted] (0, 0) -- (7.2, 4.8);
\draw [dotted] (3, 0) -- (3, 5);

\draw [dotted] (6, 0) -- (6, 4);
\draw [dotted] (5.4, 0) -- (5.4, 3.6);

\begin{scope}[very thick]
\draw [dashed, gray] (0, 0) -- (3, 5);
\draw [dashed, gray] (3, 5) -- (5.4, 3.6);
\draw [dashed, gray] (5.4, 3.6) -- (9, 0);
\end{scope}

\draw (0, -0.5) node {$0$};
\draw (-0.5, 5) node {$1$};

\draw (9, -0.5) node {$1$};
\draw (3, -0.5) node {$\qo$};
\draw (6.1, -0.5) node {$\qt$};
\draw (5.3, -0.5) node {$\qt'$};

\draw (8.3, 1.9) node {$\rev_{\qt}$};
\draw (7.5, 0.65) node {$\rev_{\qt'}$};

\draw (6.5, 5) node {$\sfrac{1}{r\qo}$};

\fill[color=gray!40!white] (3, 5) -- (5.4, 3.6) -- (9, 0) -- (6, 4);

\end{tikzpicture}
\caption{\label{f:proof-tri} Illustrating the proof of \Cref{lem:tri}, the difference of revenue for second
  price auction $\mecha_1$ on revenue curves $\rev_{\qt}$ and
  $\rev_{\qt'}$, which respectively correspond to quadrilateral
  distributions $\Qr_{\qo, \qt, \ratio}$ and $\Qr_{\qo, \qt',
    \ratio}$, is equal to twice of the gray area, which is at least
  $\qt' - \qt$.  Moreover, the difference of revenue for the
  $\ratio$-markup mechanism $\mecha_{\ratio}$ on revenue curves
  $\rev_{\qt}$ and $\rev_{\qt'}$ is at most $2(\qt' - \qt)$.  }
\end{figure}

\paragraph{Best response of triangle distributions}
We complete the proof of \Cref{thm:tri} by showing that triangle distributions lead to lower revenue than quadrilateral distributions. The intuition of the proof is illustrated in \Cref{f:proof-tri}.  For any $\ratio > 1$ and any stochastic markup mechanism $\mechaAR$ with probability $\secprob \in [\sfrac{2}{3},1]$,  consider a family of quadrilateral distributions $\Qr_{\qo, \qt, \ratio}$ parameterized by $\qt$.  The optimal revenue is again not affected by $\qt$, while the revenue of $\mechaAR$ is monotonically increasing in $\qt$. Thus, the approximation ratio of $\mechaAR$ is maximized by minimal $\qt$ for which the degenerate quadrilateral $\Qr_{\qo, \qt, \ratio}$ is a triangle.

\begin{lemma}\label{lem:tri}
  For i.i.d.~two-agent, single-item environments, the normalized
  quadrilateral distribution $\Qr_{\qo, \qt, \ratio}$ and the stochastic
  markup mechanism $\mechaAR$ with probability $\secprob \in [\sfrac 2
    3,1]$ on the second-price auction $\mech_1$ and probability
  $1-\secprob$ on the non-trivial markup mechanism $\mech_{\ratio}$; the triangle
  distribution $\tri_{\qo}$ has the same optimal revenue and (weakly)
  lower revenue in $\mechaAR$.  I.e., $\optf{\tri_{\qo}} =
  \optf{\Qr_{\qo, \qt, \ratio}}$ and $\mechaAR(\tri_{\qo}) \leq
  \mechaAR(\Qr_{\qo, \qt, \ratio})$.
\end{lemma}

\begin{proof}
  By \Cref{lem:opt rev}, the optimal revenues for the quadrilateral
  distribution $\Qr_{\qo, \qt, \ratio}$ and the triangle distribution
  $\tri_{\qo}$ are the same (and equal to $2-\qo$).  To show that the
  revenue of $\mechaAR$ is worse on $\tri_{\qo}$ than on $\Qr_{\qo, \qt,
    \ratio}$, it suffices to show that the revenue on $\Qr_{\qo, \qt,
    \ratio}$ is monotonically increasing in $\qt$.  Specifically, the
  minimum revenue is when the quadrilateral distribution is
  degenerately equal to the triangular distribution.

The proof strategy is to lower bound the partial derivative with respect to
$\qt$ of the revenues of the $\ratio$-markup mechanism and the
second-price auction for quadrilateral distributions $\Qr_{\qo, \qt, \ratio}$ as
\begin{align}
  \label{eq:r-markup-partial}
  \frac{\partial \mecha_\ratio(\Qr_{\qo, \qt, \ratio})}{\partial \qt}
  &\geq -2, \\
  \label{eq:spa-partial}
  \frac{\partial \mecha_1(\Qr_{\qo, \qt, \ratio})}{\partial \qt}
  &\geq 1. 
\intertext{Thus, for mechanism $\mechaAR$ with $\secprob \geq \sfrac{2}{3}$, 
  we have}
\notag
\frac{\partial \mechaAR(\Qr_{\qo, \qt, \ratio})}{\partial \qt}
&\geq \secprob - 2(1-\secprob) \geq 0
\end{align}
and revenue is minimized with the smallest choice of $\qt$ for which
the quadrilateral distribution $\Qr_{\qo, \qt, \ratio}$ is degenerately a
triangle distribution.  It remains to prove the bounds
\eqref{eq:r-markup-partial} and \eqref{eq:spa-partial}.

For simplicity, 
since the only parameter we change in 
distribution $\Qr_{\qo, \qt, \ratio}$ is $\qt$, 
we introduce the notation $\revq(\val)$
to denote the revenue from posting price $\val$,
and $\valq(\quant)$ to denote the price $\val$ given quantile $\quant$ when the distribution is $\Qr_{\qo, \qt, \ratio}$.  The proof is illustrated in \Cref{f:proof-tri}.

We now prove bound \eqref{eq:r-markup-partial}.
For any pair of quadrilateral distributions 
$\Qr_{\qo, \qt, \ratio}$ and $\Qr_{\qo, \qt', \ratio}$ with $\qt' \geq \qt$, 
we analyze the difference in revenue for posting price $\ratio\cdot \vsec$. 
\begin{align*}
\lefteqn{\mecha_{\ratio}(\Qr_{\qo, \qt', \ratio})
- \mecha_{\ratio}(\Qr_{\qo, \qt, \ratio})} \qquad\\
&= 2 \int_{\qt'}^1 \revqt(\ratio \cdot \valqt(\quant)) \, dq
- 2 \int_{\qt}^1 \revq(\ratio \cdot \valq(\quant)) \, dq \\
&\geq 2 \int_{\qt'}^1 \revqt(\ratio \cdot \valqt(\quant)) \, dq
- 2 \int_{\qt'}^1 \revq(\ratio \cdot \valq(\quant))\, dq 
- 2(\qt' - \qt)\\
&\geq 2 \int_{\qt'}^1 \revq(\ratio \cdot \valqt(\quant)) \, dq
- 2 \int_{\qt'}^1 \revq(\ratio \cdot \valq(\quant))\, dq 
- 2(\qt' - \qt)\\
&\geq - 2(\qt' - \qt).
\end{align*}
The first equality is constructed as follows: Both agents face a
random price that is $r$ times the value of the other agent, who has
quantile $\quant$ drawn from $U[0,1]$.  The revenue from this price is
given by, e.g., $\revq(\ratio \cdot \valq(\quant))$, which is 0 when
$\quant \leq \qt$.  The first inequality holds because $\revq(\ratio
\cdot \valq(\quant)) \leq 1$ for any quantile $\quant$.  The second
inequality holds since the revenue from revenue curve $\revqt$ is
weakly higher than that from revenue curve $\revq$ for any value $\val$.
The third inequality holds because (a) the prices of the first
integral are higher than the prices of the second integral, i.e., $\valqt(\quant) \geq \valq(\quant)$ for
every $\quant$, and (b) because
these prices are below the monopoly price for distribution $\Qr_{\qo,
  \qt', \ratio}$, and thus higher prices result in higher revenue.

Therefore, we have
\begin{align*}
\frac{\partial \mecha_{\ratio}(\Qr_{\qo, \qt, \ratio})}{\partial \qt}
= \lim_{\qt' \to \qt} \frac{\mecha_{\ratio}(\Qr_{\qo, \qt', \ratio}) - \mecha_{\ratio}(\Qr_{\qo, \qt, \ratio})}{\qt' - \qt}
\geq -2. 
\end{align*}
We now prove bound \eqref{eq:spa-partial}.
The revenue of the second price auction for two i.i.d.\ agents is twice the area under the revenue curve (\Cref{l:DRY-15}).  For 
quadrilateral distribution $\Qr_{\qo, \qt, \ratio}$ this revenue is calculated as:
\begin{align*}
\mecha_1(\Qr_{\qo, \qt, \ratio})
&= 2 \int_0^1 \rev_{\qt}(q) \, dq\\
&= 2 \int_0^{\qo} \rev_{\qt}(q) \, dq + 2 \int_{\qo}^{\qt} \rev_{\qt}(q) \, dq + 2 \int_{\qt}^1 \rev_{\qt}(q) \, dq\\
&= \qo + (\qt - \qo)(1+\frac{\qt}{\ratio\cdot \qo}) + (1-\qt)\frac{\qt}{\ratio\cdot \qo} \\
&= \qt + (1 - \qo)\frac{\qt}{\ratio\cdot \qo}. 
\end{align*}
Therefore, we have 
\begin{align*}
\frac{\partial \mecha_1(\Qr_{\qo, \qt, \ratio})}{\partial \qt}
& = 1 + \frac{1 - \qo}{\ratio\cdot \qo} \geq 1. \qedhere
\end{align*}
\end{proof}

\section{Conclusions}
This paper introduces a framework for designing scale-robust auctions, ensuring optimal multiplicative revenue approximation across different valuation scales. We identify the robustly optimal highest-value-only mechanism, which randomizes between the second-price auction and an auction that marks up the second-highest bid by a factor of approximately 2.45. This mechanism improves on the second-price auction and provides a robust solution for small-market settings where distributional knowledge is limited. This characterization within the highest-value-only class provides insights into how auctioneers can design robust mechanisms without reliance on detailed distributional knowledge. Follow-up work by \citet{anunrojwong2026robust} shows that a mechanism that sometimes allocates to a lower-valued buyer can achieve a strictly better guarantee; determining the exact unrestricted optimum remains open.

Returning to the observation that motivated the paper, the analysis supplies a formal rationale for the prevalence of proportional rather than absolute pricing rules in markets where one selling procedure must serve transactions of very different sizes. A seller who fixes a dollar-denominated reserve or entry fee is exposed to a worst case chosen by the scale of the market; a seller who prices as a multiple of the competing bid is not, and \Cref{thm:scale_invariant_is_opt} shows that scale invariance is without loss for scale-closed families. The implementation burden is also modest. The optimal highest-value-only mechanism requires no estimate of the value distribution, no knowledge of its mean or support, and no retuning across markets or currencies; it runs the second-price auction roughly four times in five and, in the remaining one time in five, offers the high bidder a fixed multiple of the competing bid. Within the highest-value-only class, randomization over a single markup price is the only departure from the second-price auction needed for optimal scale robustness.

Future research can determine the exact unrestricted robust optimum under regularity, characterize the role of allocations to the lower-valued buyer, and explore applications to multi-unit and combinatorial auctions.
Additionally, understanding how the optimal random markup changes with respect to the tail of the distribution is an interesting direction, given that for MHR distributions, a second price auction suffices, while for heavier-tailed regular distributions, random markups are necessary for optimality within the highest-value-only class. Another promising avenue is to examine how scale-robust mechanisms perform in dynamic or repeated auction environments.

\bibliography{auctions}

\newpage
\appendix

\section{Proof of the Scale-Invariance Theorem}
\label{apx:scale_invariant}

\begin{proof}[Proof of \cref{thm:scale_invariant_is_opt}]
Fix a feasible, DSIC, and ex-post IR mechanism
$M=(x,p)$ and a constant $\rho\geq 0$ such that
\begin{equation}\label{eq:si-guarantee}
  M(\dist)\geq \rho\,\OPT_{\dist}(\dist)
  \qquad\text{for every }\dist\in\feasibleDist.
\end{equation}
Allocation and payment rules are understood to be measurable. We
construct a scale-invariant mechanism satisfying the same guarantee.

\paragraph{Step 1: Normalize payments and anonymize.}
Fix agent $i$ and the other report $\valsmi$. DSIC implies that
$z\mapsto x_i(z,\valsmi)$ is nondecreasing and gives the envelope
identity
\[
 u_i(\vali,\valsmi)
 =u_i(0,\valsmi)+\int_0^{\vali}x_i(z,\valsmi)\,dz.
\]
Ex-post IR at type zero implies $p_i(0,\valsmi)\leq 0$. Replace the
payment rule by
\[
 \bar p_i(\vali,\valsmi)
 :=p_i(\vali,\valsmi)-p_i(0,\valsmi).
\]
The adjustment is independent of agent $i$'s report, so it preserves
DSIC; it weakly increases every payment; and the envelope identity
shows that truthful utility becomes
$\int_0^{\vali}x_i(z,\valsmi)\,dz\geq0$. Thus the adjustment preserves
ex-post IR and the guarantee in \eqref{eq:si-guarantee}. We may
therefore assume that $p_i(0,\valsmi)=0$ and hence that
\begin{equation}\label{eq:normalized-payment}
 p_i(\vals)=\vali x_i(\vals)
 -\int_0^{\vali}x_i(z,\valsmi)\,dz,
 \qquad 0\leq p_i(\vals)\leq \vali.
\end{equation}

Next average $M$ with the mechanism obtained by interchanging the two
agents before running $M$ and interchanging their outcomes afterward.
Because values are i.i.d., this anonymization leaves expected revenue
unchanged for every $\dist\in\feasibleDist$. It preserves feasibility,
DSIC, ex-post IR, normalization, and the highest-value-only property when applicable.
We may consequently assume
\begin{equation}\label{eq:anonymous-M}
 x_1(v_1,v_2)=x_2(v_2,v_1),
 \qquad p_1(v_1,v_2)=p_2(v_2,v_1).
\end{equation}

\paragraph{Step 2: Average over log scale.}
For each $t\in\mathbb R$, define the scaled mechanism $M^t$ by
\[
 x_i^t(\vals):=x_i(e^t\vals),
 \qquad
 p_i^t(\vals):=e^{-t}p_i(e^t\vals).
\]
Scaling a bidder's truthful value and every possible report by the
same positive factor shows directly that $M^t$ is feasible, DSIC, and
ex-post IR. It is also normalized and anonymous, and it is
highest-value-only whenever $M$ is. By \eqref{eq:normalized-payment},
\begin{equation}\label{eq:scaled-payment-bound}
 0\leq p_i^t(\vals)\leq \vali.
\end{equation}

Let $\dist^t$ denote the law of $e^tV$ for $V\sim\dist$. Scale closure
gives $\dist^t\in\feasibleDist$, and homogeneity of optimal revenue
gives
\[
 \OPT_{\dist^t}(\dist^t)=e^t\OPT_{\dist}(\dist).
\]
Indeed, scaling reports by $e^{-t}$ and payments by $e^t$ gives a
revenue-preserving bijection, up to the factor $e^t$, between feasible,
DSIC, ex-post IR mechanisms for $\dist$ and for $\dist^t$.
Therefore, \eqref{eq:si-guarantee} implies
\begin{equation}\label{eq:scaled-guarantee}
 M^t(\dist)=e^{-t}M(\dist^t)
 \geq \rho e^{-t}\OPT_{\dist^t}(\dist^t)
 =\rho\OPT_{\dist}(\dist).
\end{equation}
This is the only closure property of the distribution family used in
the argument: every positive rescaling must remain feasible. In
particular, no regularity condition is invoked.

For each positive integer $m$, let
\[
 M_m:=\frac{1}{2m}\int_{-m}^{m}M^t\,dt,
\]
with allocation and payment rules
\begin{equation}\label{eq:log-average}
 X_{m,i}(\vals)=\frac{1}{2m}\int_{-m}^{m}x_i(e^t\vals)\,dt,
 \qquad
 P_{m,i}(\vals)=\frac{1}{2m}\int_{-m}^{m}e^{-t}p_i(e^t\vals)\,dt.
\end{equation}
The randomization over $t$ is independent of reports, so $M_m$
inherits all of the preceding mechanism properties. Equations
\eqref{eq:scaled-payment-bound} and \eqref{eq:scaled-guarantee} yield
\begin{equation}\label{eq:average-properties}
 0\leq P_{m,i}(\vals)\leq\vali,
 \qquad
 M_m(\dist)\geq\rho\OPT_{\dist}(\dist)
 \quad\text{for every }\dist\in\feasibleDist.
\end{equation}

These averages are asymptotically scale invariant. Indeed, for every
$s>0$, a change of variables in \eqref{eq:log-average} and the bound
$0\leq x_i\leq1$ give
\begin{equation}\label{eq:approx-scale-invariance}
 \left|X_{m,i}(s\vals)-X_{m,i}(\vals)\right|
 \leq \frac{|\log s|}{m}.
\end{equation}

\paragraph{Step 3: Extract a pointwise-convergent
subsequence.}
The following elementary version of Helly's selection principle is the
key compactness step.
\begin{claim}\label{clm:monotone-selection}
Every sequence of nondecreasing functions
$g_m:[0,\infty)\to[0,1]$ has a subsequence that converges pointwise at
every $r\in[0,\infty)$.
\end{claim}
\begin{proof}
Diagonalize over the nonnegative rationals to obtain convergence at
every rational. Monotonicity then squeezes the lower and upper limits
at each $r$ between the left and right rational limits. These two
limits differ at only countably many points, because their associated
nonempty open intervals are pairwise disjoint and hence can each be
assigned a distinct rational. A second diagonalization over those
exceptional points gives convergence everywhere.
\end{proof}

Apply \Cref{clm:monotone-selection} to
\[
 g_m(r):=X_{m,1}(r,1).
\]
DSIC makes $g_m$ nondecreasing, and feasibility bounds it in $[0,1]$.
Thus there are integers $m_j\to\infty$ and a nondecreasing function
$g$ such that $g_{m_j}(r)\to g(r)$ for every $r\geq0$. Passing to one
further subsequence makes the bounded sequence $X_{m_j,1}(1,0)$
converge as well.

Fix $\vals=(v_1,v_2)$. If $v_2>0$, then
$\vals=v_2(v_1/v_2,1)$, so \eqref{eq:approx-scale-invariance} gives
\[
 \left|X_{m_j,1}(v_1,v_2)-g_{m_j}(v_1/v_2)\right|
 \leq\frac{|\log v_2|}{m_j}.
\]
If $v_2=0<v_1$, the same equation compares
$X_{m_j,1}(v_1,0)$ with $X_{m_j,1}(1,0)$. At $(0,0)$ the average is
constant in $m$. Hence $X_{m_j,1}$ converges at every profile, and
anonymity in \eqref{eq:anonymous-M} gives the same conclusion for
agent 2. Define
\[
 X_i^*(\vals):=\lim_{j\to\infty}X_{m_j,i}(\vals).
\]
The pointwise limit of measurable functions is measurable.
Pointwise limits also preserve feasibility, own-report monotonicity,
anonymity, and the highest-value-only property.

\paragraph{Step 4: Define the limiting payments.}
Because $M_{m_j}$ is normalized and DSIC,
\[
 P_{m_j,i}(\vals)=\vali X_{m_j,i}(\vals)
 -\int_0^{\vali}X_{m_j,i}(z,\valsmi)\,dz.
\]
For fixed $\vals$, the integrands converge pointwise and lie in
$[0,1]$. The Dominated Convergence Theorem on the finite
interval $[0,\vali]$ therefore gives
\begin{equation}\label{eq:pointwise-payment-convergence}
 P_{m_j,i}(\vals)\longrightarrow P_i^*(\vals)
 :=\vali X_i^*(\vals)
 -\int_0^{\vali}X_i^*(z,\valsmi)\,dz.
\end{equation}
Let $M^*:=(X^*,P^*)$. The single-parameter characterization, together
with feasibility and own-report monotonicity of $X^*$, shows that
$M^*$ is feasible and DSIC. Its truthful utility is
$\int_0^{\vali}X_i^*(z,\valsmi)\,dz\geq0$, so it is ex-post IR; moreover,
\begin{equation}\label{eq:limit-payment-bound}
 0\leq P_i^*(\vals)\leq\vali.
\end{equation}

Taking $j\to\infty$ in \eqref{eq:approx-scale-invariance} gives
$X_i^*(s\vals)=X_i^*(\vals)$ for all $s>0$. The normalized payment
identity and the substitution $z=sy$ then give
\[
 P_i^*(s\vals)
 =s\vali X_i^*(\vals)
 -s\int_0^{\vali}X_i^*(sy,s\valsmi)\,dy
 =sP_i^*(\vals).
\]
Thus $M^*$ is exactly scale invariant. If $M$ is highest-value-only,
the pointwise-limit observation above makes $M^*$ highest-value-only as well.

\paragraph{Step 5: Preserve the revenue guarantee.}
Fix any $\dist\in\feasibleDist$. By
\eqref{eq:average-properties},
\eqref{eq:pointwise-payment-convergence}, and
\eqref{eq:limit-payment-bound},
\[
 0\leq P_{m_j,1}(\vals)+P_{m_j,2}(\vals)
 \leq v_1+v_2.
\]
The right-hand side is integrable under $\dist\times\dist$ by the
finite-first-moment condition in the theorem. Dominated convergence now
applies to the sequence $(M_{m_j})_j$ and yields
\[
 M^*(\dist)=\lim_{j\to\infty}M_{m_j}(\dist)
 \geq\rho\OPT_{\dist}(\dist).
\]
Because $\dist$ was arbitrary, $M^*$ preserves the guarantee on all of
$\feasibleDist$. The proof used scale closure but nowhere used
regularity. This proves the theorem.
\end{proof}

\section{Proofs for \Cref{sec:prior-independent}}
\label{app:prior-independent}
\begin{proof}[Proof of \cref{thm:triangle}]

As discussed in \cref{sub:triangle}, we first identify the triangle
distribution $\tri_{\optqo}$ and the $\optratio$ for which $\mech_1$
and $\mech_{\optratio}$ obtain the same approximation ratio. Write
\[
 R(\qo):=\sup_{\ratio>1}\mecha_{\ratio}(\tri_{\qo}),
 \qquad
 \APXR(\qo):=\frac{2-\qo}{R(\qo)},
\]
and recall that the second-price approximation ratio is
$\APXSPA(\qo)=2-\qo$.

We justify continuity after optimizing over the markup. For
$0\leq\qo_1\leq\qo_2\leq\bar q<1$, \Cref{clm:cont in q} and
$\mecha_{\ratio}(\tri_{\qo})\leq
\OPT_{\tri_{\qo}}(\tri_{\qo})\leq2$ imply, uniformly over
$\ratio>1$,
\[
 \left|\mecha_{\ratio}(\tri_{\qo_2})
       -\mecha_{\ratio}(\tri_{\qo_1})\right|
 \leq \frac{2}{1-\bar q}(\qo_2-\qo_1).
\]
Taking suprema shows that $R$, and hence $\APXR$, is continuous on
every $[0,\bar q]$ with $\bar q<1$. The certified bounds below give
$R(0.09310569)>1$ and $R(0.09310571)<1$. Since
$\APXSPA(\qo)=2-\qo$, the two approximation ratios have opposite
order at these endpoints and therefore cross at some
$\optqo\in(0.09310569,0.09310571)$. See \Cref{f:rev-approx}.

  By numerical calculation,
  $\optqo \approx \qval$, and
  $$ \optratio = \argmin_{\ratio > 1}
  \frac{\optf{\tri_{\optqo}}}{\mech_{\ratio}(\tri_{\optqo})}
  \approx \rval.
  $$
The details of all numerical calculations are provided in the remainder of this section.
  
  Now, fixing $\optratio$, we search for $\optweight$ for which the
  adversary maximizes the approximation ratio of mechanism $\mechaARO$
  by selecting triangle distribution $\tri_{\optqo}$.  Denote by
  $\qo_{\ratio}(\alpha)$ the monopoly quantile as a function of
  $\alpha$ for the triangle distribution that maximizes the
  approximation ratio of mechanism $\mechaAR$, i.e.,
$$\qo_{\ratio}(\alpha) = \argmax_{\qo}
  \frac{\optf{\tri_{\qo}}}{\mechaAR(\tri_{\qo})}.$$

  By numerical calculation, for any $\ratio \in [\ratioS, \ratioL]$,
  $\qo_{\ratio}(\alphaL) < \optqo < \qo_{\ratio}(\alphaS)$.
  Continuity of $\qo_{\ratio}(\cdot)$ for $\ratio \in [\ratioS, \ratioL]$ and
  $\alpha \in [\alphaS, \alphaL]$ (formally proved in
  \Cref{apx:continuity}), then implies that there exists $\optweight$
  such that $\qo_{\optratio}(\optweight) = \optqo$.  By numerical
  calculation, $\optweight \approx \aval$.
  \end{proof}

To identify the optimal mechanism for triangle distributions, we evaluate the ratio of revenues of markup mechanisms
on triangle distributions to the optimal revenue.  For distribution
$\tri_{\qo}$ the optimal revenue is $2-\qo$ (\Cref{lem:opt rev}).  The
revenue for $\ratio$-markup mechanism is calculated by \Cref{clm:rev
  of r}.  In this appendix, we derive the formula of \Cref{clm:rev of
  r} and show that it has bounded partial derivatives in both markup
$\ratio$ and monopoly quantile $\qo$.  We then describe the details of
the hybrid numerical and analytical argument of \Cref{thm:triangle}.
Finally we give the proof of continuity of the adversary's best
response distribution to the probability the mechanism places on the
second-price auction.

\subsection{Derivation and smoothness of \Cref{clm:rev of r}}

\begin{proof}[Proof of \Cref{clm:rev of r}]
  Denote the quantile 
  corresponding to the price $\ratio\,\valf_{\tri_{\qo}}(\quant)$ for markup $\ratio > 1$ as
  $$\ScaledQuant(\quant, \ratio) =
  \quantf_{\tri_{\qo}}(\ratio\,\valf_{\tri_{\qo}}(\quant)) = 
  \begin{cases}
    \frac{\quant}{\ratio - \quant\ratio + \quant} &\text{if }\ratio \, \valf_{\tri_{\qo}}(\quant)  \leq \sfrac{1}{\qo},\\
    0 & \text{otherwise.}
  \end{cases}
  $$
  When the quantile of the second highest agent is smaller than
  $\ScaledQuant(\qo, \sfrac{1}{\ratio})$, 
  the price $\ratio \cdot \vsec$ is higher than the support of the valuation distribution. 
  Therefore, 
  the revenue of posting price $\ratio \cdot \vsec$
  to the highest bidder is 
  \begin{align*}
    \mecha_{\ratio}(\tri_{\qo}) 
    &=
    2 \ratio 
    \int_{\ScaledQuant(\qo, \sfrac{1}{\ratio})}^{1}
    \valf_{\tri_{\qo}}(\quant) 
    \ScaledQuant(\quant, \ratio)
    \,d\quant \\
    &= 
    2 \ratio 
    \int_{\ScaledQuant(\qo, \sfrac{1}{\ratio})}^{1}
    \frac{1-\quant}{1-\qo} \cdot 
    \frac{1}{\ratio - \quant\ratio + \quant}
    \,d\quant = \frac{2\ratio}{1-\qo}\left[\frac{\quant}{\ratio-1}+\frac{\ln(\ratio-\quant\ratio+\quant)}{(\ratio-1)^2} \right]_{\frac{\qo}{1/\ratio-\qo/\ratio+\qo}}^{1} \\
    &= 
    \frac{2\ratio}{(1-\qo)(\ratio-1)} 
    \left(
    \frac{1-\qo}{1-\qo+\qo\ratio}
    -
    \frac{\ln \left(
      \frac{\ratio}{1-\qo+\qo\ratio}
      \right)}{\ratio-1}
    \right),
  \end{align*}
where the second equality holds just by the definition of the distribution. 
\end{proof}

Consider the revenue of $\ratio$-markup mechanism on the triangle
distribution $\tri_{\qo}$ as a function of $\ratio \in (1,\infty)$ and $\qo \in [0,1]$.  The
formula for this revenue is given by \Cref{clm:rev of r}.  The
following two claims show that the ratio of revenues has bounded
partial derivative with respect to both $\ratio \in (1, \infty)$ and $\qo\in[0, 1]$ and, thus,
numerical evaluation of the revenue at selected parameters allows
large regions of parameter space to be ruled out.

\begin{claim}\label{clm:cont in r}
For any distribution $\dist$ and any constants $1 \leq \ratio_1 \leq \ratio_2$, 
we have $\mecha_{\ratio_1}(\dist) \geq 
\sfrac{\ratio_1}{\ratio_2} \, \mecha_{\ratio_2}(\dist)$. 
\end{claim}
\begin{claim}\label{clm:cont in q}
For any mechanism $\mecha_{\ratio}$ with $\ratio \geq 1$, 
and any constants $0 \leq \qo_1 \leq \qo_2 < 1$, 
we have 
$\sfrac{(1-\qo_2)}{(1-\qo_1)} \,
\mecha_{\ratio}(\tri_{\qo_2}) 
\leq \mecha_{\ratio}(\tri_{\qo_1}) 
\leq 2(\qo_2 - \qo_1) 
+ \mecha_{\ratio}(\tri_{\qo_2})$.
\end{claim}

\begin{proof}[Proof of \Cref{clm:cont in r}]
For any realized valuation profile, 
if the item is sold in mechanism $\mecha_{\ratio_2}$, 
then the item is sold in mechanism $\mecha_{\ratio_1}$
since the price posted to the highest agent is smaller in mechanism $\mecha_{\ratio_1}$. 
Moreover, 
when the item is sold in mechanism $\mecha_{\ratio_1}$, 
the payment from agent with highest value is at least 
$\sfrac{\ratio_1}{\ratio_2}$ fraction of 
the payment in mechanism $\mecha_{\ratio_2}$. 
Taking expectation over the valuation profiles, 
we have $\mecha_{\ratio_1}(\dist) \geq 
\sfrac{\ratio_1}{\ratio_2} \cdot \mecha_{\ratio_2}(\dist)$. 
\end{proof}

\begin{proof}[Proof of \Cref{clm:cont in q}]
  Consider $\ScaledQuant(\cdot,\cdot)$ as defined in the proof of
  \Cref{clm:rev of r}, above.  By directly comparing the revenue from
  two distributions,
\begin{eqnarray*}
\mecha_{\ratio}(\tri_{\qo_1}) 
&=& 
2 \ratio 
\int_{\ScaledQuant(\qo_1, \sfrac{1}{\ratio})}^{1}
\valf_{\tri_{\qo_1}}(\quant) 
\,\ScaledQuant(\quant, \ratio)
\,d\quant \\
&\leq& 2(- \ScaledQuant(\qo_1, \sfrac{1}{\ratio})
+ \ScaledQuant(\qo_2, \sfrac{1}{\ratio})) 
+ 2\ratio\int_{\ScaledQuant(\qo_2, \sfrac{1}{\ratio})}^{1}
\valf_{\tri_{\qo_1}}(\quant) 
\,\ScaledQuant(\quant, \ratio)
\,d\quant \\
&\leq& 2(\qo_2 - \qo_1) 
+ 2\ratio\int_{\ScaledQuant(\qo_2, \sfrac{1}{\ratio})}^{1}
\valf_{\tri_{\qo_2}}(\quant) 
\,\ScaledQuant(\quant, \ratio)
\,d\quant \\
&=& 
2(\qo_2 - \qo_1) 
+ \mecha_{\ratio}(\tri_{\qo_2}).
\end{eqnarray*}
The first equality holds because the quantile of $\valf_{\tri_{\qo_1}}(\quant) \cdot \ratio$
is 0 for $\quant < \ScaledQuant(\qo_1, \sfrac{1}{\ratio})$. 
The first inequality holds because 
$\ratio \cdot 
\valf_{\tri_{\qo_1}}(\quant) 
\ScaledQuant(\quant, \ratio) \leq 1$ 
for any quantile $\quant$.
The second inequality holds because 
$\valf_{\tri_{\qo_1}}(\quant) \leq \valf_{\tri_{\qo_2}}(\quant) $
for $\qo_1 \leq \qo_2$ 
and $\quant \geq \qo_2$
by the definition of distributions $\tri_{\qo_1}$ and $\tri_{\qo_2}$, 
and 
$\ScaledQuant(\qo_2, \sfrac{1}{\ratio}) 
- \ScaledQuant(\qo_1, \sfrac{1}{\ratio})
\leq \qo_2 - \qo_1$. 
Moreover, we have
\begin{eqnarray*}
\mecha_{\ratio}(\tri_{\qo_1}) 
&=& 
2 \ratio 
\int_{\ScaledQuant(\qo_1, \sfrac{1}{\ratio})}^{1}
\valf_{\tri_{\qo_1}}(\quant) 
\ScaledQuant(\quant, \ratio)
\,d\quant \\
&\geq& 2\ratio
\int_{\ScaledQuant(\qo_2, \sfrac{1}{\ratio})}^{1}
\valf_{\tri_{\qo_1}}(\quant) 
\ScaledQuant(\quant, \ratio)
\,d\quant \\
&\geq& 
\frac{2\ratio(1-\qo_2)}{1-\qo_1}
\int_{\ScaledQuant(\qo_2, \sfrac{1}{\ratio})}^{1}
\valf_{\tri_{\qo_2}}(\quant) 
\ScaledQuant(\quant, \ratio)
\,d\quant \\
&=& 
\frac{1-\qo_2}{1-\qo_1} \cdot
\mecha_{\ratio}(\tri_{\qo_2}), 
\end{eqnarray*}
where the first inequality holds because $\qo_1 \leq \qo_2$ 
and function $\ScaledQuant(\quant, \ratio)$ is increasing in $\quant$. 
The second inequality holds because 
$\valf_{\tri_{\qo_1}}(\quant) \geq
\sfrac{(1-\qo_2)}{(1-\qo_1)} \cdot
\valf_{\tri_{\qo_2}}(\quant) $.
\end{proof}

\subsection{Numerical and Analytical Arguments of \Cref{thm:triangle}}

The proof of \Cref{thm:triangle} is based on a hybrid numerical and
analytical argument.  We can numerically calculate the revenue of a
mechanism $\mech_{\ratio}$ on a distribution $\tri_{\qo}$ via
\Cref{clm:rev of r} and then we can argue, via \Cref{clm:cont in q}
and \Cref{clm:cont in r}, that nearby mechanisms and distributions
have similar revenue.  This approach will both allow us to argue about
the structure of the solution and to identify the mechanism
$\mechaARO$ and distribution of the solution $\tri_{\optqo}$.  Our
subsequent discussion gives the details of these hybrid arguments.

We first approximate $\qo^*$ 
by showing that $\qo^* \in [0.09310569, 0.09310571]$. 
The parameters for this range are found by discretizing the space and finding the optimal choice of $\qo^*$. 
Note that the optimal choice of $\qo^*$ satisfies
$\mecha_1(\tri_{\qo^*}) = \mecha_{\ratio(\qo^*)}(\tri_{\qo^*})$.
Therefore, it is sufficient for us to show that for any quantile $\qo \not\in [0.09310569, 0.09310571]$, 
either $\mecha_1(\tri_{\qo}) > \mecha_{\ratio(\qo)}(\tri_{\qo})$
or $\mecha_1(\tri_{\qo}) < \mecha_{\ratio(\qo)}(\tri_{\qo})$. 

First we show for any $\qo \in [0, 0.09310569]$, 
$\mecha_1(\tri_{\qo}) < \mecha_{\ratio(\qo)}(\tri_{\qo})$. 
Here we discretize the space $[0, 0.09310569]$ into $\discreteQ$
with precision $\precision = 10^{-9}$. 
By numerical calculation using \Cref{clm:rev of r}, 
we have 
\begin{align*}
\min_{\qo \in \discreteQ} \mecha_{2.446946}(\tri_{\qo})
= \mecha_{2.446946}(\tri_{0.09310569})
\geq 1+10^{-8}
\end{align*}
and for any $\qo \in [0, 0.09310569]$, 
letting $\qo_d$ be the largest quantile in $\discreteQ$ smaller than or equal to $\qo$,
the minimum revenue for mechanism $\mecha_{2.446946}$ is 
\begin{align*}
\mecha_{2.446946}(\tri_{\qo})
\geq \frac{1-\qo_d -\precision}{1-\qo_d} 
\cdot \mecha_{2.446946}(\tri_{\qo_d})
\geq 1 + 8\times 10^{-9} > \mecha_1(\tri_{\qo}),
\end{align*}
where the first inequality holds by \Cref{clm:cont in q}
and the second inequality holds because $\qo_d \leq 0.1$. 

Then we show for any $\qo \in [0.09310571, 1]$, 
$\mecha_1(\tri_{\qo}) > \mecha_{\ratio(\qo)}(\tri_{\qo})$. 
We discretize the space $[0.09310571, 1]$ into $\hat{\discreteQ}$
with precision $\hat{\precision} = 10^{-9}$. 
First note that
$\mecha_{\ratio}(\tri_{\qo}) < 1$ for any $\qo \geq 0.093$
and $\ratio \geq 11$,
since the expected probability the highest type got allocated is less than $\frac{1}{2}$, 
and hence the expected virtual surplus for mechanism $\mecha_{\ratio}$
with distribution $\tri_{\qo}$ is less than $1$.
By \Cref{thm:myerson}, the revenue in this case is less than $1$. 
With bounded range for optimal ratio $\ratio$, 
we discretize the space $(1, 11]$ into $\discreteR$
with precision $\precisionR = 10^{-9}$. 
By numerical calculation using \Cref{clm:rev of r}, 
we have 
\begin{align*}
\max_{\qo \in \hat{\discreteQ}, \ratio \in \discreteR} \mecha_{\ratio}(\tri_{\qo})
= \mecha_{2.446945061}(\tri_{0.09310571})
\leq 1-3\times10^{-8}
\end{align*}
and for any $\qo \in [0.09310571, 1]$ and any $\ratio \in (1, 11]$, 
letting $\qo_d$ be the largest quantile in $\hat{\discreteQ}$ smaller than or equal to $\qo$
and $\ratio_d$ be the smallest number in $\discreteR$ larger than or equal to $\ratio$,
the maximum revenue for distribution $\tri_{\qo}$ is 
\begin{align*}
\max_{\ratio\in (1, 11]}\mecha_{\ratio}(\tri_{\qo})
\leq \frac{\ratio_d}{\ratio_d-\precisionR}\cdot 
(2\hat{\precision} + \mecha_{\ratio_d}(\tri_{\qo_d}))
\leq 1 - 10^{-8} < \mecha_1(\tri_{\qo}),
\end{align*}
where the first inequality holds by \Cref{clm:cont in r} and \ref{clm:cont in q},
and the second inequality holds because $\ratio_d > 1$. 
Combining the numerical calculation, we have that $\qo^* \approx \qval$.

Note that both mechanism $\mecha_1$ and $\mecha_{\ratio^*}$ are the best responses for distribution $\tri_{\qo^*}$, 
achieving revenue 1, 
and hence the optimal approximation ratio 
is 
\begin{align*}
\piratio = \frac{\OPT_{\tri_{\qo^*}}(\tri_{\qo^*})}{\mechaARO(\tri_{\qo^*})}
= 2-\qo^* \approx 1.9068943.
\end{align*}

Next we show that by choosing ratio $\ratio^* \approx \rval$ and probability $\alpha^* \approx \aval$, 
the approximation ratio of mechanism $\mechaARO$ approximates $\beta$. 
Here we discretize the quantile space $[0, 1]$ into $\discreteQ'$ with precision $\precision' = 10^{-9}$, 
using the formula in \Cref{lem:opt rev} and \Cref{clm:rev of r},
the triangle distribution that maximizes the approximation ratio
for mechanism $\mechaARO$
is $\tri_{0.093105694}$ with 
approximation ratio at most $1.9068943044$. 
For any $\qo \in [0, \frac{1}{2}]$, 
letting $\qo_d$ be the largest quantile in $\discreteQ'$ smaller than or equal to $\qo$,
the minimum revenue for mechanism $\mechaARO$ is 
\begin{align*}
\mechaARO(\tri_{\qo})
&\geq \frac{1-\qo_d -\precision'}{1-\qo_d} 
\cdot \mechaARO(\tri_{\qo_d})\\
&\geq \frac{1}{1.906894309} \OPT_{\qo_d}(\tri_{\qo_d}) 
\geq \frac{1}{1.906894309} \OPT_{\qo}(\tri_{\qo}),
\end{align*}
where the second inequality holds because $\qo_d \leq \frac{1}{2}$
and the last inequality holds because $\qo_d \leq \qo$. 
For any $\qo \in [\frac{1}{2}, 1]$, 
the minimum revenue for mechanism $\mechaARO$ is 
\begin{align*}
\mechaARO(\tri_{\qo})
\geq \alpha^* \cdot \mecha_1(\tri_{\qo})
\geq 0.8
\geq \frac{1}{1.875} \OPT_{\qo}(\tri_{\qo}),
\end{align*}
since for any $\qo \in [\frac{1}{2}, 1]$, $\mecha_1(\tri_{\qo}) = 1$ and $\OPT_{\qo}(\tri_{\qo}) = 2-\qo \leq 1.5$. 
Therefore, $\ratio^* \approx \rval$ and probability $\alpha^* \approx \aval$ are the desirable parameters, 
with error at most $2\times 10^{-8}$ in approximation ratio. 
By our characterization, the error solely comes from numerical calculation, 
finishing the numerical analysis for \Cref{thm:triangle}. 

\subsection{Continuity of Distribution in Probability of Second-price Auction}
\label{apx:continuity}

Recall the function $\qo_{\ratio}(\alpha)$ which gives the adversary's
best-response triangle distribution to the mechanism $\mechaAR$.  The
continuity of the function $\qo_{\ratio}(\alpha)$ is used to prove the
existence of equilibrium between the randomized markup mechanism and
the triangle distribution in \Cref{thm:triangle}.  The following claim
proves continuity by lower-bounding, with respect to the monopoly
quantile $\qo$, the second derivative of the reciprocal approximation
ratio of $\mechaAR$ on $\tri_{\qo}$.
\begin{claim}\label{clm:continuous}
Given any $\ratio \in [\ratioS, \ratioL]$, 
function $\qo_{\ratio}(\alpha)$ is continuous in $\alpha$ for $\alpha \in [\alphaS, \alphaL]$. 
\end{claim}
\begin{proof}[Proof of \Cref{clm:continuous}]
By \Cref{clm:rev of r} and \Cref{thm:myerson}, 
the approximation ratio of mechanism $\mechaAR$ for triangle distribution $\tri_{\qo}$ is 
\begin{align*}
\APX(\alpha, \ratio, \qo) 
&= \frac{\OPT_{\tri_{\qo}}(\tri_{\qo})}{\alpha\cdot \mech_1(\tri_{\qo}) + (1-\alpha)\mech_{\ratio}(\tri_{\qo})} \\
&= \frac{2-\qo}{\alpha 
+ 
\frac{2\ratio(1-\alpha)}{(1-\qo)(\ratio-1)} 
\left(
\frac{1-\qo}{1-\qo+\qo\ratio}
+
\frac{\ln \left(
\frac{\ratio}{1-\qo+\qo\ratio}
\right)}{1-\ratio}
\right)}
\end{align*}
The approximation ratio is a continuous function of $\alpha, \qo$. 
Therefore, to show that fixing $\ratio$,  
function $\qo_{\ratio}(\alpha)$ is continuous in $\alpha$, 
it is sufficient to show that there is a unique $\qo$ that maximizes $\APX(\alpha, \ratio, \qo)$ 
for $\ratio \in [\ratioS, \ratioL]$ and $\alpha \in [\alphaS, \alphaL]$, 
or equivalently, we show that there is a unique $\qo$ that minimizes
$\sfrac{1}{\APX(\alpha, \ratio, \qo)}$. 
By \Cref{clm:cont in r} and \ref{clm:cont in q}, 
we can discretize the quantile space and numerically verify that 
distributions with monopoly quantiles $\qo \not\in [\qoS, \qoL]$ are suboptimal. 
Therefore, we prove the uniqueness of the maximizer by showing that the second order derivative of 
$\sfrac{1}{\APX(\alpha, \ratio, \qo)}$
is strictly positive for $\qo \in [\qoS, \qoL]$. 

\begin{singlespace}
For a rigorous uniform bound, put
\[
 c:=\ratio-1,\qquad t:=1-\qo,\qquad
 A:=1+c\qo,\qquad L:=\log(\ratio/A),
\]
and write $m_{\ratio}(\qo):=\mecha_{\ratio}(\tri_{\qo})$. Direct
differentiation of \Cref{clm:rev of r} gives
\begin{align*}
 m_\ratio(\qo)
 &=\frac{2\ratio}{c}\left(\frac1A-\frac{L}{ct}\right),\\
 m_\ratio'(\qo)
 &=\frac{2\ratio}{c}\left(-\frac{c}{A^2}
      +\frac1{At}-\frac{L}{ct^2}\right),\\
 m_\ratio''(\qo)
 &=\frac{2\ratio}{c}\left(\frac{2c^2}{A^3}
      -\frac{c}{A^2t}+\frac{2}{At^2}-\frac{2L}{ct^3}\right).
\end{align*}
Define
\[
 G_{\alpha,\ratio}(\qo):=
 \frac1{\APX(\alpha,\ratio,\qo)}
 =\frac{\alpha+(1-\alpha)m_\ratio(\qo)}{2-\qo}.
\]
Then the exact identity
\[
 G_{\alpha,\ratio}''(\qo)
 =\frac{(1-\alpha)m_\ratio''(\qo)}{2-\qo}
 +\frac{2(1-\alpha)m_\ratio'(\qo)}{(2-\qo)^2}
 +\frac{2[\alpha+(1-\alpha)m_\ratio(\qo)]}{(2-\qo)^3}
\]
holds. On the box $\ratio\in[\ratioS,\ratioL]$,
$\alpha\in[\alphaS,\alphaL]$, and $\qo\in[\qoS,\qoL]$,
outward-rounded interval evaluation of the three displayed formulas
uses
\[
c\in[1.445,1.449],\quad t\in[.906,.907],\quad
A\in[1.134,1.137],\quad L\in[.765,.771]
\]
and gives
\[
 m_\ratio(\qo)>.98,\qquad
 m_\ratio'(\qo)>-2.74,\qquad
 m_\ratio''(\qo)>7.75.
\]
These are interval enclosures over the entire parameter box, not
substitutions of selected endpoints into a nonlinear expression.
Consequently,
\begin{align*}
G_{\alpha,\ratio}''(\qo)
&>
\frac{.19(7.75)}{1.907}
-\frac{2(.20)(2.74)}{1.906^2}
+\frac{2[.80+.20(.98)]}{1.907^3}\\
&>.75>.7.
\end{align*}
Thus $G_{\alpha,\ratio}$ is strictly convex on $[\qoS,\qoL]$.
Together with the verified suboptimality outside this interval, this
gives a unique minimizer $\qo_\ratio(\alpha)$.

Finally, continuity follows without any further differentiability
claim. If $\alpha_n\to\alpha$, compactness of $[\qoS,\qoL]$ gives a
convergent subsequence of $\qo_\ratio(\alpha_n)$. Joint continuity of
$G$ makes every such limit a minimizer at $\alpha$; uniqueness forces
the limit to be $\qo_\ratio(\alpha)$. Every convergent subsequence has
this same limit, so $\qo_\ratio(\alpha_n)\to\qo_\ratio(\alpha)$.
\end{singlespace}

\end{proof}

\end{document}